\documentclass[12pt,aps,eqsecnum,nofootinbib,floatfix]{revtex4}
\usepackage{bbding}
\usepackage{pifont}
\usepackage{subfigure}
\usepackage{epsfig}
\usepackage{array}%\usepackage{CJK}
\usepackage{amsmath}
\usepackage{amsfonts}
\usepackage{amssymb}
\usepackage{color}
\usepackage{graphicx}
\usepackage{mathrsfs}
\usepackage{multirow}

\def\be{\begin{equation}}
\def\ee{\end{equation}}
\def\ba{\begin{eqnarray}}
\def\ea{\end{eqnarray}}

\newcommand{\omits}[1]{}

\begin{document}

\title{Clapeyron equation and phase equilibrium properties in higher dimensional charged topological dilaton AdS black holes with a nonlinear source}

\author{Huai-Fan Li\footnote{corresponding author: Email: huaifan.li@stu.xjtu.edu.cn(Huai-Fan Li)}, Hui-Hua Zhao, Li-Chun Zhang, Ren Zhao}

\medskip

\affiliation{Institute of Theoretical Physics, Shanxi Datong University,
Datong 037009, China}
\affiliation{Department of Physics, Shanxi Datong University,
Datong 037009, China}

\begin{abstract}
\begin{center}
\large{Abstract}
\end{center}
Using Maxwell's equal area law, we discuss the phase transition of higher dimensional charged topological dilaton AdS black hole with a nonlinear source. The coexisting region of the two phases is found and we depict the coexistence region in $P-v$ diagrams. The two-phase equilibrium curves in $P-T$ diagrams are plotted, and we take the first order approximation of volume $v$ in the calculation. To better compare with a general thermodynamic system, the Clapeyron equation is derived for higher dimensional charged topological black hole with a nonlinear source. The latent heat of isothermal phase transition is investigated. We also study the effect of the parameters of the black hole on the region of two-phases coexistence.  The results show that the black hole may go through a small-large phase transition similar to those of usual non-gravity thermodynamic systems.

\end{abstract}

\bigskip
\maketitle

\section{Introduction}

Black holes have been used as the laboratory of many kinds of theories, specially
the thermodynamics of black hole plays an important role. It has been found that a
black hole possesses not only standard thermodynamic quantities but also abundant phase
structures and critical phenomena, such as the Hawking-Page phase
transition~\cite{Hawking}, similar to a general thermodynamic system. Even more interesting
is that the studies on the charged black holes show that they may have an analogous phase transition
with that of van der Walls-Maxwell's liquid-gas~\cite{Cai1,David}. In recent years, the idea of including the variation of the  cosmological constant
$\Lambda$ in the first law of black hole thermodynamics has attracted increasing attention~\cite{Cai1,David,Cai,Kastor,Weinberg}. In the case of an asymptotical AdS and dS black hole in $n$ dimensions, one identifies the pressure of black hole with the cosmological constant $\Lambda$~\cite{Cai}
\begin{equation}
\label{eq01} P=-\frac{1}{8 \pi} \Lambda=\frac{n(n-1)}{16 \pi l^2}
\end{equation}
and the corresponding conjugate quantity, thermodynamic
volume~\cite{Kastor}
\begin{equation}
\label{eq02}
V=\left( {\frac{\partial M}{\partial P}} \right)_{S,Q_i ,J_k}.
\end{equation}
It may seem a little surprising to elevate $\Lambda$ to the status of a thermodynamic
variable. $\Lambda$ is usually thought as an energy in the Einstein
action. However, there are some physical reasons to view the cosmological constant as a variable of perssure. For example, one may suppose
that there exist several more fundamental theories where some physical constants such as Yukawa coupling, gauge
coupling constants, Newton's constant, and/or cosmological constant, may not be fixed values but dynamical
ones arising from the vacuum expectation values. In that case, it is natural to add these variations of "constants"
into the first law of black hole thermodynamics~\cite{David,Cai,Kastor,Weinberg}. The $(P\sim V)$ critical behaviors in AdS and dS black holes have
been extensively studied~\cite{Dola,Guna,Frass,David1,Altami,Altami1,Altami2,
Zhao,Zhao1,Zhao2,Ma,Zhang,Ma1,Ma2,Heidi,Heidi1,Heidi2,Heidi3,Heidi4,Heidi5,Heidi6,Arci,
Banerjee,Banerjee1,Banerjee2,Majhi,Majhi1,Chakraborty,Chakraborty1,Chakraborty2,Peca,Peca1, Mo2,Mo3,Mo4,Mo5,Lala,Wei,Suresh,Mans,Niu,Ma3,Zou,Zou1,Li,Wei1,Poshteh,Xu,Liu,Xu1,Wei2,Seyed}. A completely simulated gas-liquid
system has been put forward.

Using Ehrenfest scheme, Ref.\cite{Mo2,Mo3,Mo4,Mo5,Lala} studied the critical phenomena in
a series of black holes in AdS spacetime, and proved the phase transition at
critical point is second order, which has also been confirmed in
Ref.\cite{Kastor,Niu2,Ong} by studying thermodynamics and state space geometry of black
holes.  Although some encouraging results about black hole thermodynamic
properties in AdS and dS spacetimes have been achieved and the problems
about phase transition of black holes have been extensively discussed,
a unified recognition about the phase transition of black hole has not
 been put forward. It is significant to further explore phase
equilibrium and phase structure in black holes, which can
help to recognize the evolution of  black holes.

The motivation for studying higher dimensional black holes comes
from developments in string/M-theory, which is believed to be the
most consistent approach to quantum theory of gravity. It was argued that black holes may play a crucial
role in the analysis of dynamics in higher dimensions as well as
in the compactification mechanisms. In particular, to test novel
predictions of string/M-theory black holes may serve
as good theoretical laboratories. It has been thought that the
statistical-mechanical calculation of the Bekenstein-Hawking
entropy for a class of supersymmetric black holes in five dimensions
is one of the remarkable results in string theory~\cite{Stro,Brec}.
Another motivation on studying higher dimensional black holes originates
from the braneworld scenarios, as a new fundamental scale of quantum gravity.
An interesting consequence of these models is the possibility of mini black
hole production at future accelerating colliders~\cite{Dimo}. It is nontrivial to study the higher
dimensional spacetime, because the dimension of spacetime will affect the
thermodynamical properties of the black hole~\cite{Emparan,Chak,Davies,Myung}.

The theory of nonlinear electrodynamics was first introduced in 1930's by Born and Infeld to
obtain a classical theory of charged particles with finite self-energy~\cite{Born}.
Born-Infeld(BI) theory has received lots attentions since it turns out
to play an important role in string theory. The BI action, including a dilaton
and an axion field, appears in the coupling of an open superstring and an Abelian gauge
field thory~\cite{Frad}. Here we would like to consider another type of nonlinear electrodynamics,
namely, the exponential form of the nonlinear electrodynamics in the setup
of dilaton gravity. These are our main motivation to
explore the effects of dilaton field on the properties of higher dimensional charged
black holes with a nonlinear source.

The isotherms in $P\sim v$ diagrams of charged topological dilaton
AdS black hole in Ref.~\cite{Zhao1} show that there exists thermodynamic
unstable region with $\partial P/\partial v>0$ when temperature is
below critical temperature and the negative pressure emerges when
temperature is below a certain value. This situation also exists in
van der Waals-Maxwell gas-liquid system, which has been resolved by
Maxwell's equal area law. In this paper, using the Maxwell's equal area law,
we establish a phase transition process in charged topological dilaton
AdS black holes, where the issues about unstable states and negative pressure
are resolved. By studying the phase transition process, we acquire the two-phase
equilibrium properties including the $P-T$ phase diagram, Clapeyron equation and latent heat of phase change.
The results show the phase transition below critical temperature is of the first order but phase transition
at critical point belongs to the continuous one,i.e., the second order phase transition.

Outline of this paper is as follows: The higher dimensional charged topological dilaton AdS
black hole as a thermodynamic system is briefly introduced in
section 2. In section 3, by Maxwell equal area law the phase
transition processes at certain temperatures are obtained and the
boundary of two phase equilibrium region are depicted in $P-v$
diagram for a higher dimensional charged topological dilaton AdS black hole. Then
some parameters of the black hole are analyzed to find the relevance
with the two-phase equilibrium. In section 4, the $P-T$ phase diagrams are plotted
and the Clapeyron equation and latent heat of the phase change are derived.
We make some discussions and conclusions in section 5.
(we use the units $G_d =\hbar =k_B=c=1$ in this paper)

\section{The Thermodynamic quantities of Higher-Dimensional Charged Dilaton Black Holes}

We consider the $n$-dimensional$(n\geq4)$ action in which gravity is coupled to dilaton and
nonlinear electrodynamic field. The motivation for choosing the model includes: First, the
solution of this model might lead to possible extensions of AdS/CFT correspondence. Second,
such solutions of this model may be used to extend the range of validity of methods and tools originally
developed for, asymptotically flat or asymptotically (A)dS
black holes.The Einstein-Maxwell-Dilaton(EMD) action in $n$-dimensional
spacetime is~\cite{Sheykhi}
\begin{equation}
\label{eq1} S=\frac{1}{16\pi }\int {d^{n}} x\sqrt {-g} \left(
{R-\frac{4}{n-2}(\nabla \Phi )^2-V(\Phi )-L(F, \Phi)} \right),
\end{equation}
where the dilaton potential contains two Liouville terms:
\begin{equation}
\label{eq2} V(\Phi) =2\Lambda_0 e^{2 \varsigma_0 \Phi} + 2\Lambda e^{2 \varsigma \Phi},
\end{equation}
and
\begin{equation}
\label{eq021}L(F,\Phi)=4 \beta^2 e^{4\alpha \Phi/(n-2)}\left[\exp\left(-\frac{e^{-8\alpha \Phi/(n-2)}F^2}{4\beta^2}\right)-1 \right]
\end{equation}
where $R$ is the Ricci scalar curvature, $\Phi$ is the dilaton field and
$V(\Phi )$ is a potential for $\Phi $, $\Lambda _0 $, $\Lambda $, $\varsigma
_0 $ and $\varsigma $ are parameters. $\alpha $ is a parameter determining the strength of coupling of the scalar and
electromagnetic field. This kind of potential was previously
investigated in the context of BI-dilaton (BId) black holes~\cite{Ma2,Heidi} as well
as EMD gravity~\cite{Mo4,Mo5,Lala,Wei,Suresh,Mans}. In the limit $\beta \rightarrow\infty$, the lagrangian of the exponential nonlinear
electrodynamics coupled to the dilaton field recovers the standard linear Maxwell Lagrangian coupled to
the dilaton field. The obtained solutions fully satisfy
the system of eq.(\ref{eq1}) provided we take
\begin{equation}
\label{eq601}
\varsigma_0=1/\alpha,\quad \varsigma=\alpha,\quad \Lambda_0=b^{-2} \alpha^2/(\alpha^2-1).
\end{equation}
Notice that $\Lambda$ remains as a free parameter which plays the role of
the cosmological constant~\cite{Sheykhi}.

The metric of such a spacetime can be written
\begin{equation}
\label{eq3}
ds^2=-f(r)dt^2+\frac{dr^2}{f(r)}+r^2R^2(r)d\Omega _{n-2}^2.
\end{equation}
The more details of the metric function $f(r)$ can be found in the appendix~\ref{A}.
The location of black horizon satisfies the equation $f(r_+)=0$, from which we obtain
\[
m = - \frac{(n - 3)(\alpha ^2 + 1)^2b^{ - \gamma }}{(\alpha ^2 - 1)(\alpha
^2 + n - 3)}r_+^{n - 3 - (n - 4)\gamma / 2} + \frac{2\Lambda (\alpha ^2 +
1)^2b^\gamma }{(n - 2)(\alpha ^2 - n + 1)}r_+^{n - 1 - n\gamma / 2}
\]
\begin{equation}
\label{eq12}
 + \frac{2q^2(\alpha ^2 + 1)^2b^{ - (n - 3)\gamma }}{(n - 2)(\alpha ^2 + n -
3)r_+^{ - (n - 4)\gamma / 2 + n - 3}} - \frac{q^4(\alpha ^2 + 1)^2b^{ - (2n -
5)\gamma }}{2\beta ^2(n - 2)(\alpha ^2 + 3n - 7)r_+^{3n - 7 + (8 - 3n)\gamma /
2}} + o\left( {\frac{1}{\beta ^4}} \right).
\end{equation}
In (\ref{eq12}), $m$ serves as an integration constant. According to the
definition of mass due to Abbott and Deser~\cite{Abbott}, the ADM(Arnowitt-Deser-Misnsr)
mass of the solution (\ref{eq11}) is
\[
M(S,Q,P) = \frac{b^{(n - 2)\gamma / 2}(n - 2)\omega _{n - 2} }{16\pi (\alpha
^2 + 1)}m
\]

\[
 = - \frac{(n - 3)(n - 2)\omega _{n - 2} (\alpha ^2 + 1)b^{(n - 4)\gamma /
2}}{16\pi (\alpha ^2 - 1)(\alpha ^2 + n - 3)}r^{n - 3 - (n - 4)\gamma / 2} +
\frac{2\Lambda (\alpha ^2 + 1)\omega _{n - 2} b^{n\gamma / 2}}{16\pi (\alpha
^2 - n + 1)}r^{n - 1 - n\gamma / 2}
\]

\begin{equation}
\label{eq14}
 + \frac{2q^2(\alpha ^2 + 1)\omega _{n - 2} b^{ - (n - 4)\gamma / 2}}{16\pi
(\alpha ^2 + n - 3)r^{ - (n - 4)\gamma / 2 + n - 3}}
 - \frac{q^4(\alpha ^2 + 1)\omega _{n - 2} b^{ - (3n - 8)\gamma / 2}}{32\pi
\beta ^2(\alpha ^2 + 3n - 7)r^{3n - 7 + (8 - 3n)\gamma / 2}} + o\left(
{\frac{1}{\beta ^4}} \right).
\end{equation}
The entropy of the EMD black hole still satisfies the so called area law of the
entropy which states that the entropy of the black hole is a quarter of the
event horizon area~\cite{Wald}. This universal law applies to almost all kinds
of black holes, including dilaton black holes, in Einstein gravity~\cite{Barnich,Padmanabhan}. It
is easy to show
\begin{equation}
\label{eq15}
S = \frac{b^{(n - 2)\gamma / 2}\omega _{n - 2} r_ + ^{(n - 2)(1 - \gamma /
2)} }{4}.
\end{equation}
One may then regard the parameters $S$, $Q$ and $P$ as a complete set of
parameters for the mass $M(S,Q,P)$ and define the parameters
conjugate to $S$, $Q$ and $P$. These quantities are the
temperature, the electric potential and volume
\begin{equation}
\label{eq17}
T = \left( {\frac{\partial M}{\partial S}} \right)_{Q,P} ,
\quad
U = \left( {\frac{\partial M}{\partial Q}} \right)_{S,P} ,
\quad
V = \left( {\frac{\partial M}{\partial P}} \right)_{Q,S} ,
\end{equation}
where the temperature of Hawking radiation
\[
T = - \frac{(n - 3)(1 + \alpha ^2)b^{ - \gamma }}{4\pi (\alpha ^2 - 1)}r_ +
^{\gamma - 1}
 - \frac{\Lambda (1 + \alpha ^2)b^\gamma }{2\pi (n - 2)}r_ + ^{1 - \gamma }
\]
\begin{equation}
\label{eq18}
 - \frac{q^2(1 + \alpha ^2)b^{ - (n - 3)\gamma }}{2\pi (n - 2)}r_ +
^{(n\gamma + 5 - 3\gamma - 2n)}
 + \frac{q^4(1 + \alpha ^2)b^{ - (2n - 5)\gamma }}{8\pi (n - 2)\beta ^2}r_ +
^{9 + 2n\gamma - 4n - 5\gamma } + o\left( {\frac{1}{\beta ^4}} \right),
\end{equation}
the electric potential
\begin{equation}
\label{eq19}
U = \frac{4\pi Q(\alpha ^2 + 1)b^{(4 - n)\gamma / 2}}{(\alpha ^2 + n -
3)\omega _{n - 2} r_ + ^{n - 3 + 2\gamma - n\gamma / 2} } - \frac{(4\pi
)^3Q^3(\alpha ^2 + 1)b^{ - (3n - 8)\gamma / 2}}{2\beta ^2\omega _{n - 2}^3
(\alpha ^2 + 3n - 7)r^{3n - 7 + (8 - 3n)\gamma / 2}} + o\left(
{\frac{1}{\beta ^4}} \right),
\end{equation}
and the volume
\begin{equation}
\label{eq20}
V = - \frac{(\alpha ^2 + 1)b^{\gamma n / 2}\omega _{n - 2} }{(\alpha ^2 - n
+ 1)}r_ + ^{n - 1 - \gamma n / 2} .
\end{equation}
Note that in the limit $\beta \to \infty $, Eq. (\ref{eq18})
reduces to the temperature of higher dimensional EMD black holes~\cite{Lala}. Thus,
the thermodynamics quantities satisfy the first law of thermodynamics
\begin{equation}
\label{eq21}
dM = TdS + UdQ + VdP.
\end{equation}
For the electric potential satisfies the superposition principle, we can rewrite the
equation (\ref{eq19}),
\begin{equation}
\label{eq22}
U = U_1 + U_2 ,
\end{equation}
where
\[
U_1 = \frac{4\pi Q(\alpha ^2 + 1)b^{(4 - n)\gamma / 2}}{(\alpha ^2 + n -
3)\omega _{n - 2} r_ + ^{n - 3 + 2\gamma - n\gamma / 2} },
\]
\begin{equation}
\label{eq23}
U_2 = - \frac{(4\pi )^3Q^3(\alpha ^2 + 1)b^{ - (3n - 8)\gamma / 2}}{2\beta
^2\omega _{n - 2}^3 (\alpha ^2 + 3n - 7)r^{3n - 7 + (8 - 3n)\gamma / 2}} +
o\left( {\frac{1}{\beta ^4}} \right).
\end{equation}
So the Smarr formula:
\begin{equation}
\label{eq24}
M = \frac{(n - 2)}{\alpha ^2 + n - 3}TS + U_1 Q + \frac{(\alpha ^2 + 2n -
5)}{2(\alpha ^2 + n - 3)}U_2 Q + \frac{2(\alpha ^2 - 1)}{\alpha ^2 + n -
3}VP.
\end{equation}
Notice that the relations eqs.(\ref{eq21}) and (\ref{eq24}) are satisfying the given family of solutions.
We can rewrite the eq.(\ref{eq18}),
\[
P = \frac{T}{v}
 + \frac{k(n - 2)(\alpha ^2 + 1)^2}{\pi (n - 1)(\alpha ^2 - 1)v^2}
 + \frac{Q^2b^{2(1 - n)\gamma }2\pi }{\omega _{n - 1}^2 }\left( {\frac{v(n -
1)}{4(\alpha ^2 + 1)b^{2\gamma }}} \right)^{\textstyle{{(2 - n)(2 - \gamma
)} \over {1 - 2\gamma }}}
\]

\[
 - \frac{Q^4(4\pi )^3(1 + \alpha ^2)b^{ - (2n - 5)\gamma }}{2\omega _{n -
2}^4 (n - 2)\beta ^2}r_ + ^{9 + 2n\gamma - 4n - 5\gamma }
\]

\begin{equation}
\label{eq25}
 = \frac{T}{v} - \frac{A}{v^2} + \frac{BQ^2}{v^d} - \frac{CQ^4}{v^{2d}},
\end{equation}
with the specific volume~\cite{Zhao1}
\begin{equation}
\label{eq26}
v = \frac{4(\alpha ^2 + 1)b^\gamma }{(n - 2)}r_ + ^{1 - \gamma } ,
\end{equation}
and
\[
d = \frac{(n - 2)(2 - \gamma )}{1 - \gamma },
\quad
A = \frac{(n - 3)(\alpha ^2 + 1)^2}{\pi (n - 2)(1 - \alpha ^2)},
\]

\[
B = \frac{b^{(2 - n)\gamma }2\pi }{\omega _{n - 1}^2 }\left( {\frac{4(\alpha
^2 + 1)b^{2\gamma }}{(n - 2)}} \right)^{(n - 2)(2 - \gamma ) / (1 - \gamma
)},
\]

\begin{equation}
\label{eq27}
C = \frac{b^{2(2 - n)\gamma }(4\pi )^3}{8\omega _{n - 1}^4 \beta ^2}\left(
{\frac{4(\alpha ^2 + 1)b^{2\gamma }}{(n - 2)}} \right)^{2(n - 2)(2 - \gamma
) / (1 - \gamma )}
\end{equation}
In Fig.1 we plot the isotherms in $P-v$ diagrams in terms of state equation Eq.
(\ref{eq25}) at different dimension $n$, charge $Q$, and parameters
$b$ and $\alpha $. One can see from Fig.1 that there are
thermodynamic unstable regimes with $\partial P/\partial v>0$ on
the isotherms when temperature $T<T_c $, where $T_c $ is critical
temperature. When the temperature $T=T_0$, there is a point of
intersection between the isotherms and the horizontal $v$ axis.
 And the negative pressure emerges when temperature is
below certain value $T_0$. $T_0$ and the corresponding
specific volume $v_0$ can be derived
\begin{equation}
\label{eq20}
Av_0^{2d - 1} = BQ^2(d - 1)v_0^d - CQ^4(2d - 1),
\quad
T_0 = \frac{A}{v_0 } - \frac{BQ^2}{v_0^{d - 1} } + \frac{CQ^4}{v_0^{2d - 1}
}.
\end{equation}

\begin{figure}[!htbp]
\center{\subfigure[~$n=4$,$b=0.8$,$Q=1.5$,$\alpha=0$] {
\includegraphics[angle=0,width=5cm,keepaspectratio]{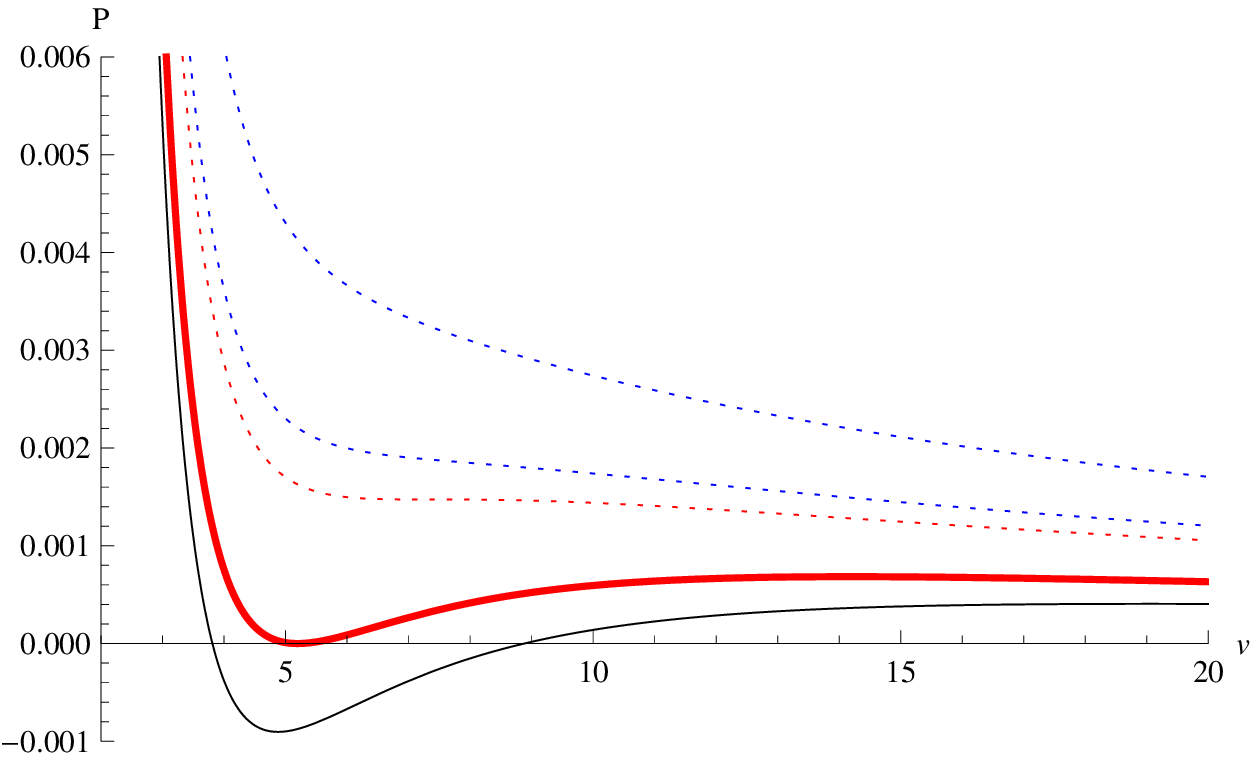}}
\subfigure[~$n=5$,$b=1$,$Q=1.2$,$\alpha=\frac{1}{\sqrt{2}}$] {
\includegraphics[angle=0,width=5cm,keepaspectratio]{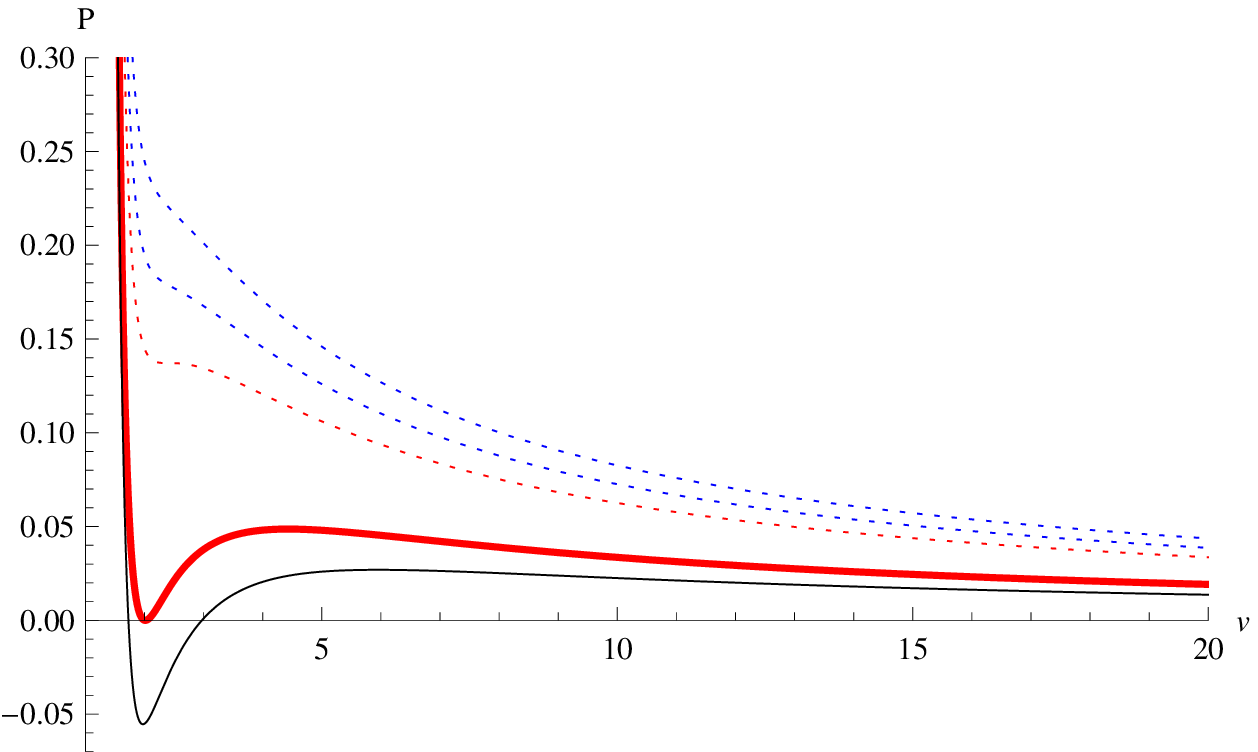}}
\subfigure[~$n=6$,$b=1.2$,$Q=1$,$\alpha=\frac{1}{\sqrt{3}}$] {
\includegraphics[angle=0,width=5cm,keepaspectratio]{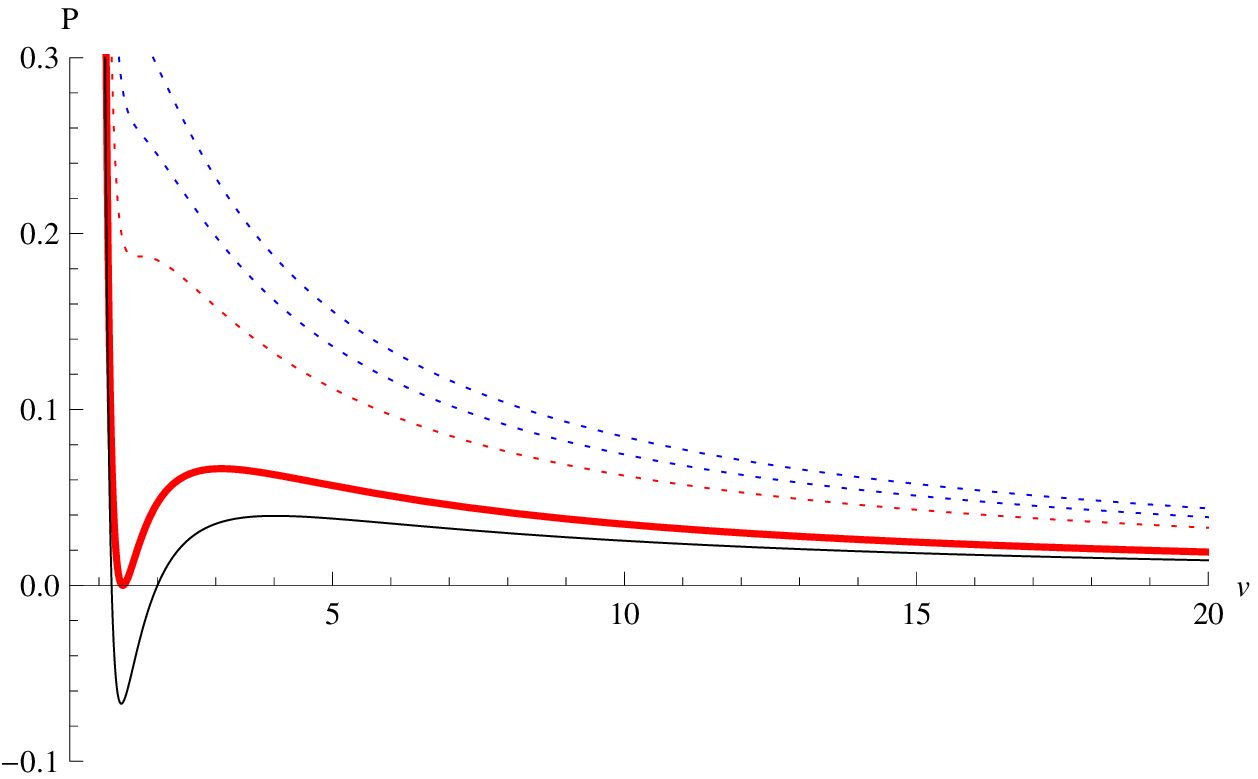}}
\caption[]{\it Isotherms in $P-v$ diagrams of higher-dimensional charged topological dilaton black holes
in $n$ dimensional AdS spacetime}} \label{Pv1}
\end{figure}

\section{two-Phase equilibrium and Maxwell's equal area law}

The equation of state of the charged topological black hole is exhibited by the
isotherms in Fig.1, in which the thermodynamic unstable states
with $\partial P/\partial v>0$ will lead to the system
automatically expand or contract unrestrictedly. The cases occur also in van der
Waals equation but they have been resolved by Maxwell's equal area law.

We extend the Maxwell's equal area law to $n$-dimensional charged
topological dilaton AdS black hole to establish an phase transition process of the black
hole as a thermodynamic system. On the isotherm with temperature $T_0 $ in $P-v$
diagram, the two points $\left( {P_0 ,\;v_1 } \right)$ and $\left( {P_0
,\;v_2 } \right)$ meet the Maxwell's equal area law,
\begin{equation}
\label{eq28}
P_0 (v_2 -v_1 )=\int\limits_{v_1 }^{v_2 } {Pdv} ,
\end{equation}
which results in
\[
P_0 (v_2 - v_1 ) = T_0 \ln \left( {\frac{v_2 }{v_1 }} \right) - A\left(
{\frac{1}{v_1 } - \frac{1}{v_2 }} \right)
\]
\begin{equation}
\label{eq29}
 + \frac{BQ^2}{d - 1}\left( {\frac{1}{v_1^{d - 1} } - \frac{1}{v_2^{d - 1}
}} \right) - \frac{CQ^4}{2d - 1}\left( {\frac{1}{v_1^{2d - 1} } -
\frac{1}{v_2^{2d - 1} }} \right),
\end{equation}
where the two points $\left( {P_0 ,\;v_1 } \right)$ and $\left( {P_0 ,\;v_2
} \right)$ are seen as endpoints of isothermal phase transition. Considering
\begin{equation}
\label{eq30}
P_0 = \frac{T_0 }{v_1 } - \frac{A}{v_1^2 } + \frac{BQ^2}{v_1^d } -
\frac{CQ^4}{v_1^{2d} },
\quad
P_0 = \frac{T_0 }{v_2 } - \frac{A}{v_2^2 } + \frac{B}{v_2^d } -
\frac{CQ^4}{v_2^{2d} },
\end{equation}
we can get
\begin{equation}
\label{eq31}
0 = T_0 \left( {\frac{1}{v_1 } - \frac{1}{v_2 }} \right) - A\left(
{\frac{1}{v_1^2 } - \frac{1}{v_2^2 }} \right) + BQ^2\left( {\frac{1}{v_1^d }
- \frac{1}{v_2^d }} \right) - CQ^4\left( {\frac{1}{v_1^{2d} } -
\frac{1}{v_2^{2d} }} \right),
\end{equation}

\begin{equation}
\label{eq32}
2P_0 = T_0 \left( {\frac{1}{v_1 } + \frac{1}{v_2 }} \right) - A\left(
{\frac{1}{v_1^2 } + \frac{1}{v_2^2 }} \right) + BQ^2\left( {\frac{1}{v_1^d }
+ \frac{1}{v_2^d }} \right) - CQ^4\left( {\frac{1}{v_1^{2d} } +
\frac{1}{v_2^{2d} }} \right),
\end{equation}
Using the eqs.(\ref{eq29}),(\ref{eq30}) and (\ref{eq31}), we can get
\[
Av_2^{2d - 2} x^{2d - 2}\left( {2(1 - x) + (1 + x)\ln x} \right)
 = BQ^2v_2^d x^d\left( {\frac{d(1 - x^{d - 1})}{(d - 1)} + \frac{(1 -
x^d)\ln x}{(1 - x)}} \right)
\]

\begin{equation}
\label{eq33}
 - CQ^4\left( {\frac{2d(1 - x^{2d - 1})}{(2d - 1)} + \frac{(1 - x^{2d})\ln
x}{(1 - x)}} \right),
\end{equation}
with $x=v_1 /v_2 $ $(0<x<1)$. When $x$ is given, we can obtain the $v_2$ and $v_1$ corresponding
to the $x$ from (\ref{eq33}).

For $C$ is little, taking the zeroth order approximation, we can obtain
\begin{equation}
\label{eq34}
v_{2;0}^{d - 2} = \frac{B}{A}\frac{d(1 - x^{d - 1})(1 - x) + (d - 1)(1 -
x^d)\ln x}{x^{d - 2}(d - 1)(1 - x)\left( {2(1 - x) + (1 + x)\ln x} \right)}
= f^{d - 2}(x).
\end{equation}
So, we can obtain the first order approximation solution $v_{2;1}^{d - 2}$ by substituting the $v_{2;0}^{d - 2}$
to the eq.(\ref{eq33}). Similarly, we can obtain the second order approximation solution  $v_{2;2}^{d - 2}$ by
substituting the $v_{2;2}^{d - 2}$ to the eq.(\ref{eq33}). And so on, we can obtain the arbitrary order
approximation solution. For convenience, we take
\begin{equation}
\label{eq35}
v_2 = F(x).
\end{equation}
In the first order approximation, we have
\[
v_{2;1}^{2d - 2} = \left( {\frac{d(1 - x^{d - 1})}{(d - 1)} + \frac{(1 -
x^d)\ln x}{(1 - x)}} \right)\frac{BQ^2f^{d}(x)}{Ax^{d - 2}\left(
{2(1 - x) + (1 + x)\ln x} \right)}
\]

\begin{equation}
\label{eq36}
 - \frac{CQ^4}{Ax^{2d - 2}\left( {2(1 - x) + (1 + x)\ln x} \right)}\left(
{\frac{2d(1 - x^{2d - 1})}{(2d - 1)} + \frac{(1 - x^{2d})\ln x}{(1 - x)}}
\right) \approx F^{2d - 2}(x).
\end{equation}
When $x\to 1$, the corresponding state is critical point state. From (\ref{eq34}) and (\ref{eq36}), we can obtain
\[
v_c^{2d - 2}
 = v_c^d \mathop {\lim }\limits_{x \to 1} \frac{BQ^2}{Ax^{d - 2}\left( {2(1
- x) + (1 + x)\ln x} \right)}\left( {\frac{d(1 - x^{d - 1})}{(d - 1)} +
\frac{(1 - x^d)\ln x}{(1 - x)}} \right)
\]

\begin{equation}
\label{eq37}
 - \mathop {\lim }\limits_{x \to 1} \frac{CQ^4}{Ax^{2d - 2}\left( {2(1 - x)
+ (1 + x)\ln x} \right)}\left( {\frac{2d(1 - x^{2d - 1})}{(2d - 1)} +
\frac{(1 - x^{2d})\ln x}{(1 - x)}} \right).
\end{equation}
From (\ref{eq31}), we obtain
\begin{equation}
\label{eq38}
Tv_2^{2d - 1} x^{2d - 1} = Av_2^{2d - 2} x^{2d - 2}(1 + x) - BQ^2\frac{v_2^d
x^d(1 - x^d)}{1 - x}
 + CQ^4\frac{(1 - x^{2d})}{1 - x}.
\end{equation}
Substituting (\ref{eq35}) into (\ref{eq38}), we can obtain
\begin{equation}
\label{eq39}
\chi T_c x^{2d - 1}F^{2d - 1}(x) = Ax^{2d - 2}F^{2d - 2}(x)(1 + x) -
BQ^2\frac{x^dF^d(x)(1 - x^d)}{1 - x}
 + CQ^4\frac{(1 - x^{2d})}{1 - x},
\end{equation}
with $T=\chi T_c$. Because we take account of the case that the temperature $T$ below the critical temperature $T_c$
, the value of $\chi=\frac{T}{T_c}$ is from $0$ to $1$. When $x\to 1$ and $\chi \to 1$ , the corresponding state
is critical state. For a fixed $\chi $, i.e. a fixed $T_0 $, we can get a certain $x$ from Eq. (\ref{eq39}),
and then according to Eqs. (\ref{eq30}) and (\ref{eq33}), the $v_2 $ and $P_0
$ are solved.

\begin{figure}
  \centering
  \includegraphics[width=4in]{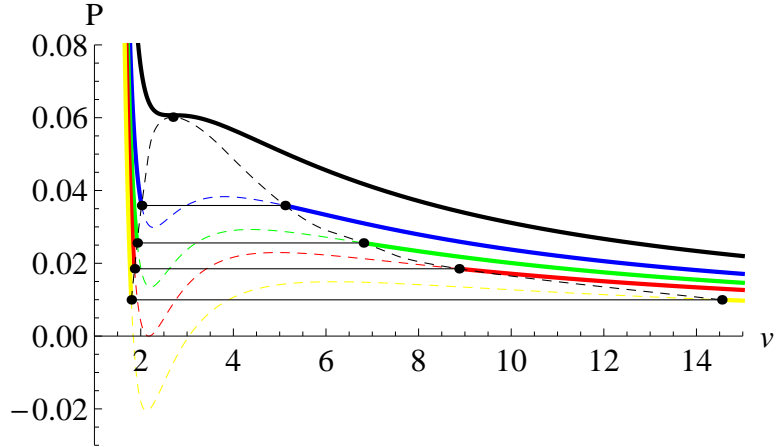}\\
  \caption{\it The simulated isothermal phase transition by isobars and the boundary
of two phase coexistence region for the topological dilaton black hole as
$n=5$, $b=1.2$, $Q=1.2$, $\alpha =\frac{1}{\sqrt{3}}$.}\label{coexist}
\end{figure}

\section{Two-phase coexistent curves and the phase change latent}

Due to the lack of knowledge of chemical potential, the $P-T$ curves of two-phase
equilibrium coexistence for general thermodynamic system are usually
obtained by experiment. However the slope of the curves can be calculated
by Clapeyron equation in theory,
\begin{equation}
\label{eq40}
\frac{dP}{dT}=\frac{L}{T(v^\beta -v^\alpha )},
\end{equation}
where the latent heat of phase change  $L=T(s^\beta -s^\alpha )$, $v^\alpha $,
$s^\alpha $ and $v^\beta $, $s^\beta $ are the molar volumes and molar
entropy of phase $\alpha $ and phase $\beta $ respectively. So Clapeyron equation
provides a direct experimental verification for some phase transition theories.

Here we investigate the two phase equilibrium coexistence $P-T$ curves and
the slope of them for the topological dilaton AdS black hole. From the (\ref{eq39}), we can obtain
the temperature
\[
T = A\frac{(1 + x)}{xF(x)} - BQ^2\frac{(1 - x^d)}{(1 - x)x^{d - 1}F^{d -
1}(x)}
 + CQ^4\frac{(1 - x^{2d})}{1 - x}
\]

\begin{equation}
\label{eq41}
 + CQ^4\frac{(1 - x^{2d})}{(1 - x)x^{2d - 1}F^{2d - 1}(x)} = G(x),
\end{equation}
Meanwhile, from the the eq.(\ref{eq25}), we obtain
\begin{equation}
\label{eq42}
P = \frac{T}{F(x)}
 - \frac{A}{F^2(x)}
 + \frac{BQ^2}{F^d(x)}
 - \frac{CQ^4}{F^{2d}(x)},
\end{equation}
Substituting the eq.(\ref{eq41}) to eq.(\ref{eq42}), we can obtain
\begin{equation}
\label{eq43}
P = \frac{G(x)}{F(x)} - \frac{A}{F^2(x)} + \frac{BQ^2}{F^d(x)} - \frac{CQ^4}{F^{2d}(x)} = H(x).
\end{equation}

\begin{figure}
  \centering
  \includegraphics[angle=0,width=5cm,keepaspectratio]{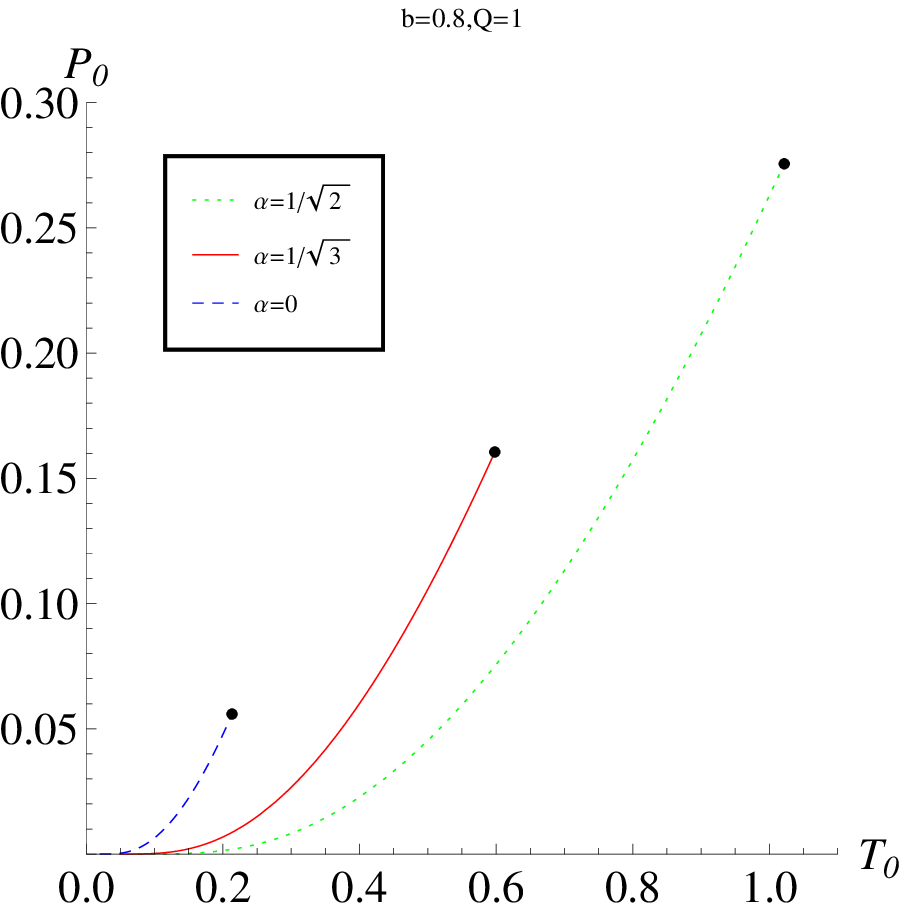}
  \includegraphics[angle=0,width=5cm,keepaspectratio]{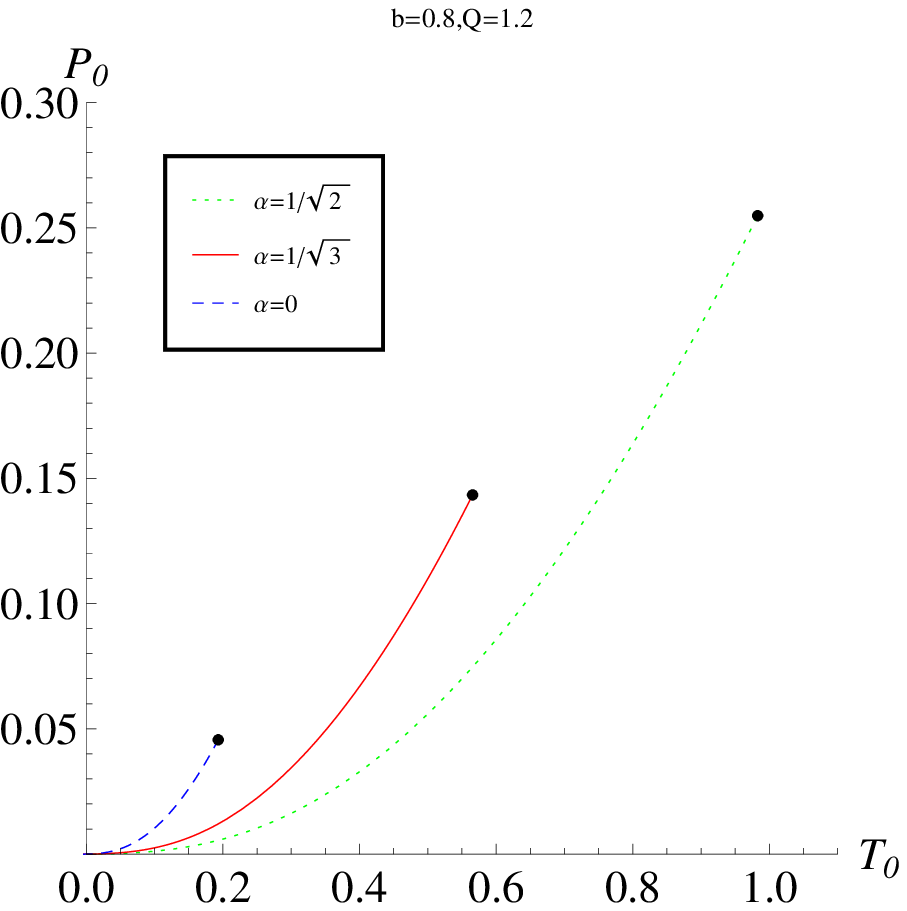}
  \includegraphics[angle=0,width=5cm,keepaspectratio]{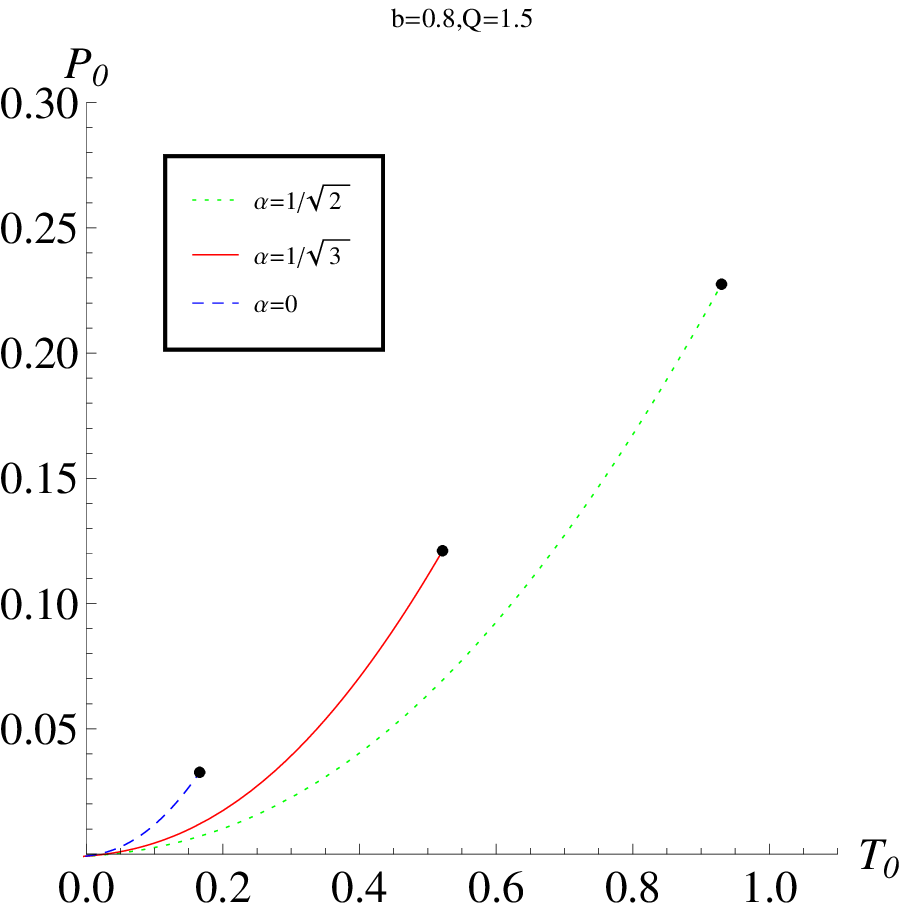}
  \includegraphics[angle=0,width=5cm,keepaspectratio]{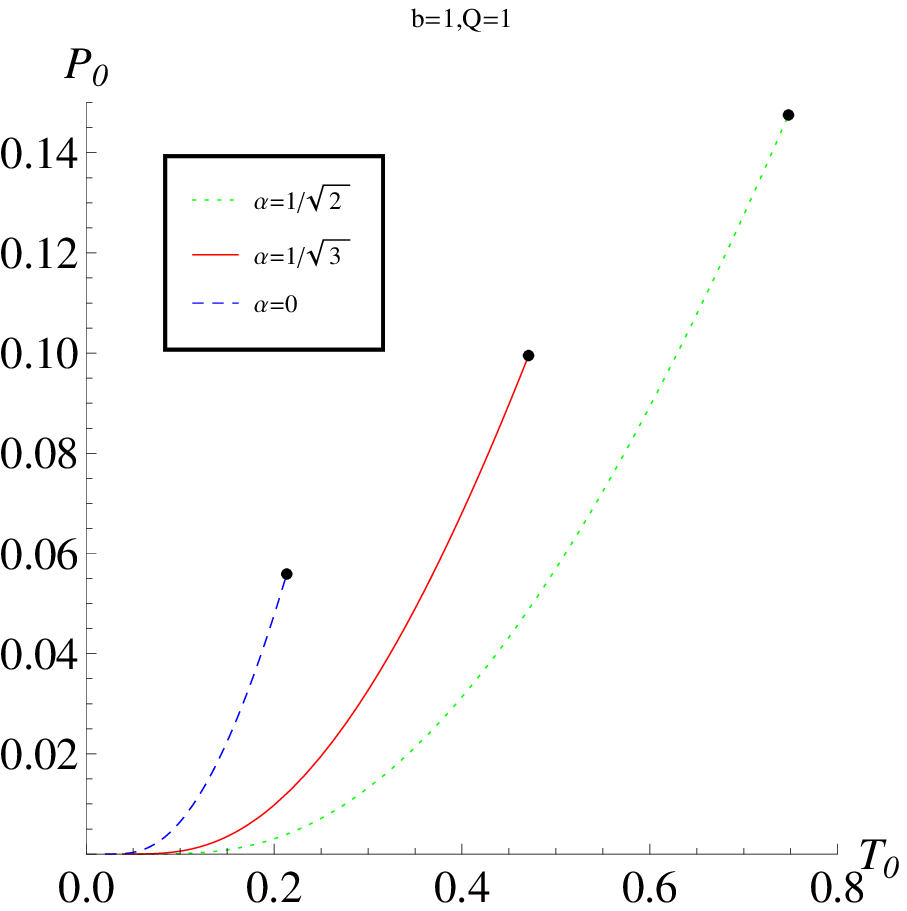}
  \includegraphics[angle=0,width=5cm,keepaspectratio]{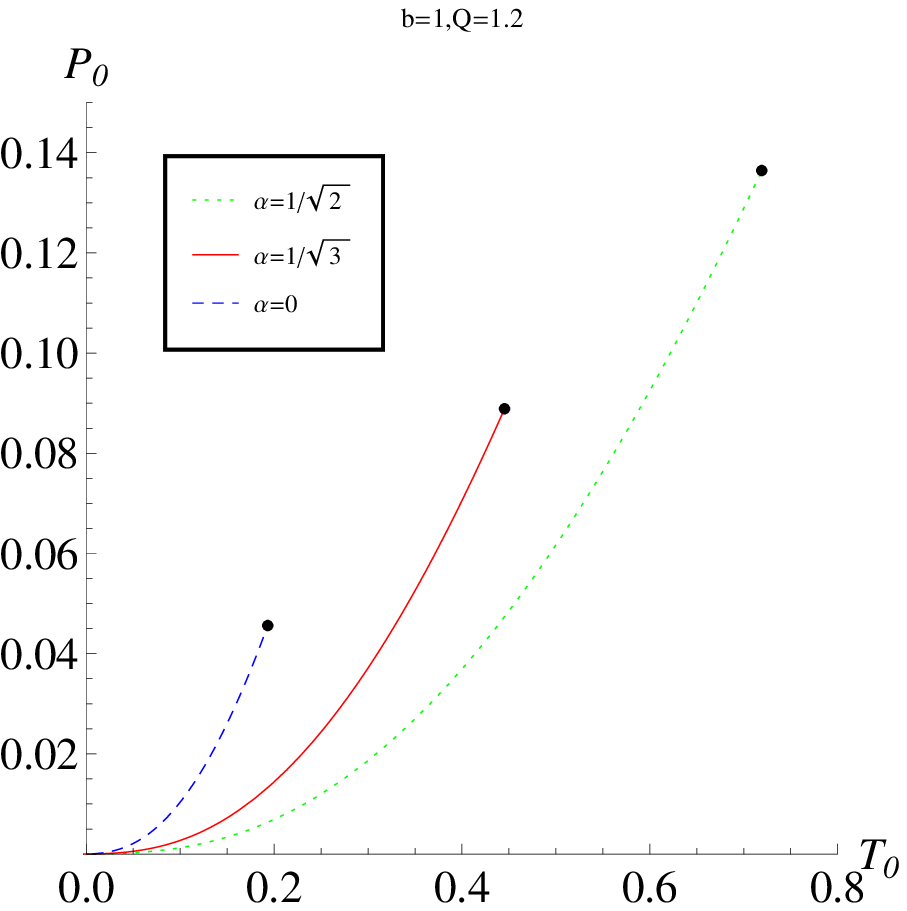}
  \includegraphics[angle=0,width=5cm,keepaspectratio]{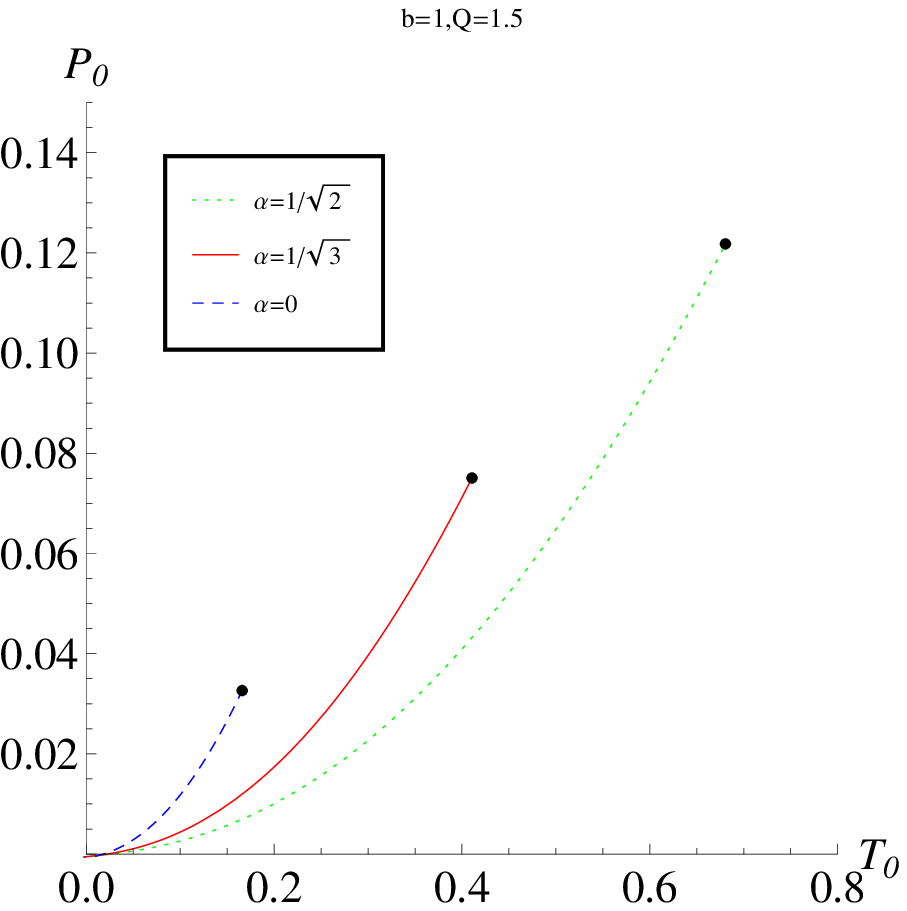}
  \includegraphics[angle=0,width=5cm,keepaspectratio]{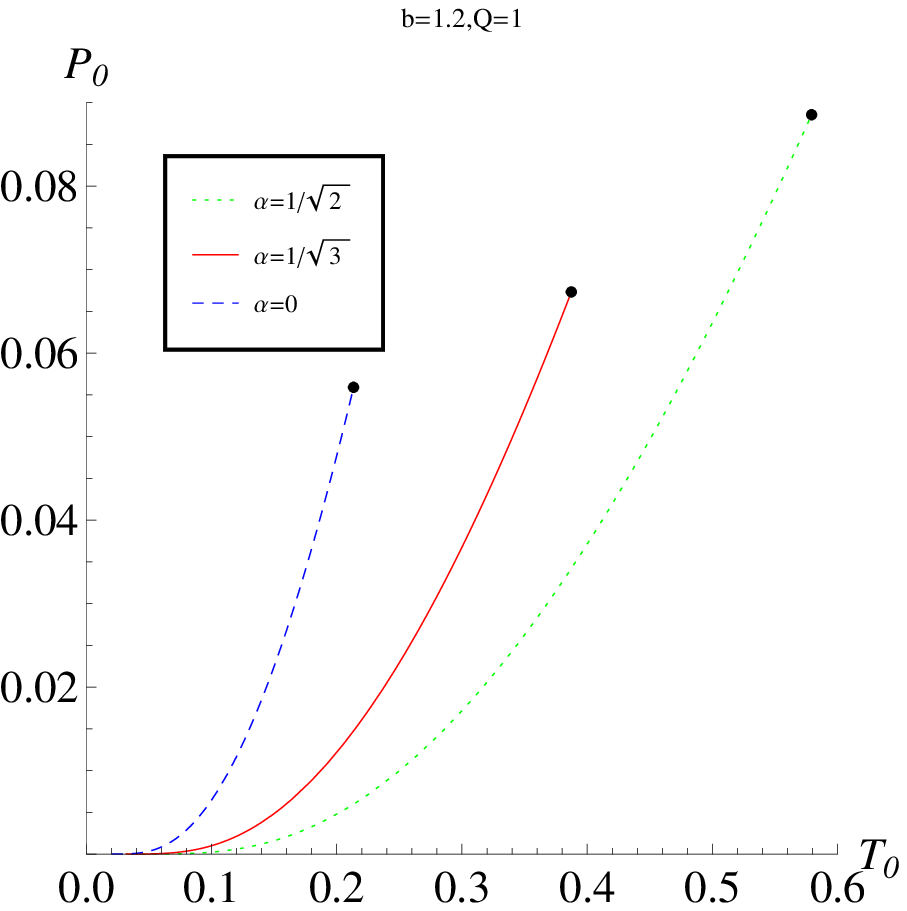}
  \includegraphics[angle=0,width=5cm,keepaspectratio]{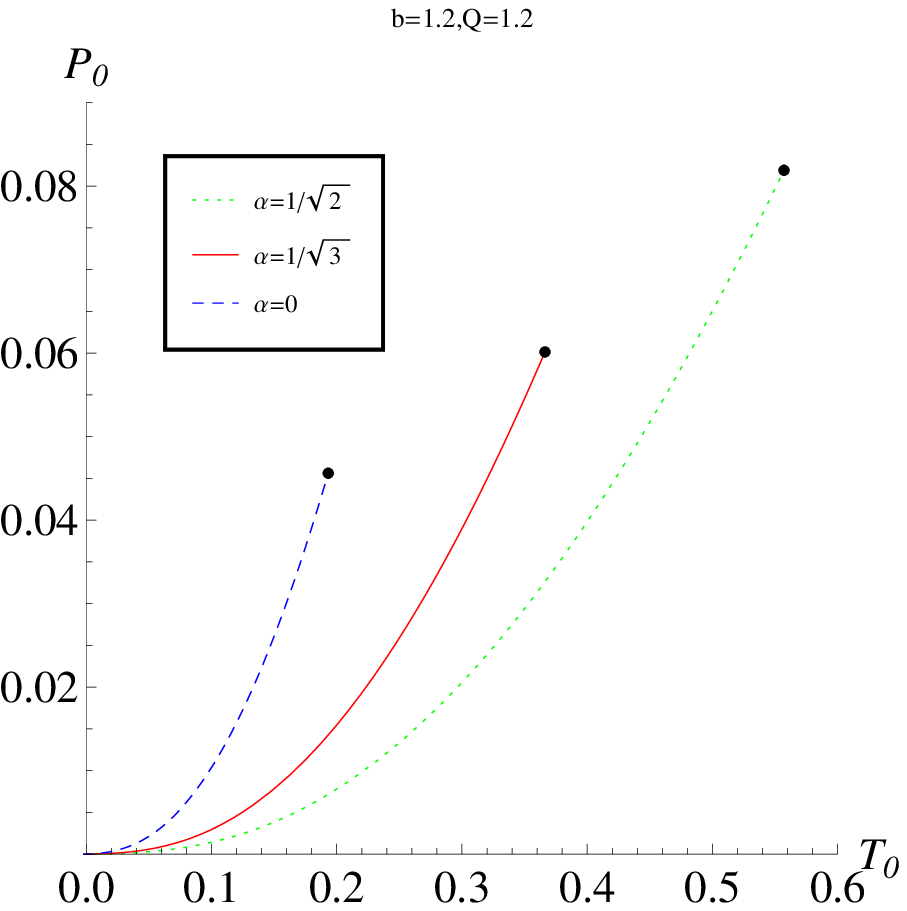}
  \includegraphics[angle=0,width=5cm,keepaspectratio]{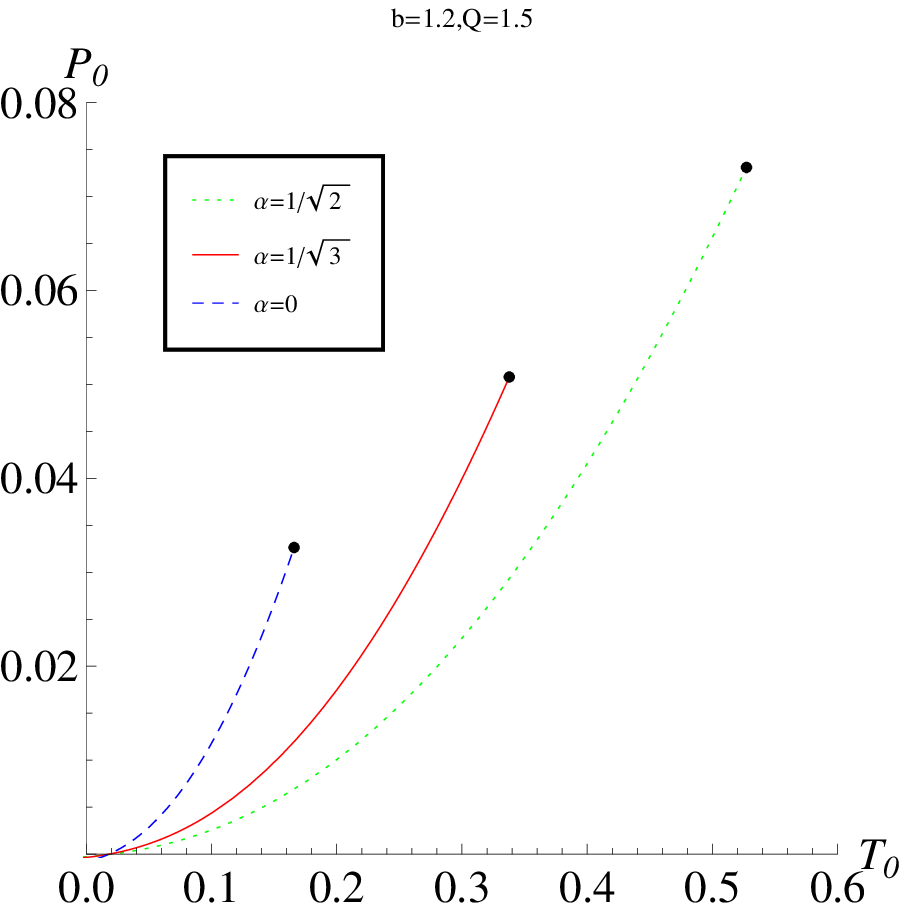}\\
  \caption{\it Two phase equilibrium coexistence curves in $P-T$ diagrams for the
topological dilaton black hole in $5$-dimensional AdS spacetime. In
the diagram, the first line with $b=0.8$, the Second line with $b=1.0$, the final line with $b=1.2$ and the first column with
$Q=1.0$, the second column with $Q=1.2$, the final column with $Q=1.5$}\label{P0-T0}
\end{figure}
We plot the $P-T$ curves with $0<x\le 1$ in Fig.3 when the parameter
$b$, coupling parameter $\alpha$, charge $Q$ take different values respectively. The curves
represent two-phase equilibrium condition for the
topological dilaton AdS black hole and the terminal points of the
curves represent corresponding critical points. Fig.3 shows that for fixed parameter $b$ and charge $Q$, both
the critical temperature and pressure increases as coupling parameter $\alpha$ increases.
Both critical temperature and pressure increases with coupling parameter $\alpha$, but two-phase equilibrium pressure decreases with increasing parameter $b$ at certain
temperature. The change of two-phase equilibrium curve
with charge $Q$ is similar to that with parameter $b$. As charge $Q$
becomes larger the critical pressure and critical temperature become smaller.

From Eq.(\ref{eq43}), we obtain
\begin{equation}
\label{eq44}
\frac{dP}{dT}=\frac{H '(x)}{G'(x)},
\end{equation}
where $H'(x)=\frac{dH}{dx}$. The Eq. (\ref{eq44}) represents the
slope of two-phase equilibrium $P-T$ curve as a function of $x$.
From Eqs.(\ref{eq40}) and (\ref{eq44}), we can get the latent heat of phase change
as function of $x$ for $n$-dimensional charged topological dilaton AdS black hole,
\begin{equation}
\label{eq45}
L=T(1-x)\frac{H '(x)}{G '(x)}F(x)=(1-x)\frac{H '(x)}{G
'(x)}G (x)F(x).
\end{equation}
We plot the $L-x$ curves with $0<x\le 1$ in Fig.4, as
the parameter $b$, coupling parameter $\alpha $ and charge $Q$ take certain values. From
Fig.5 we can see that the effects of $x$ and the spacetime dimension $n
$ and charge $Q$ on phase change latent heat $L$.
\begin{figure}
  \centering
  \includegraphics[angle=0,width=5cm,keepaspectratio]{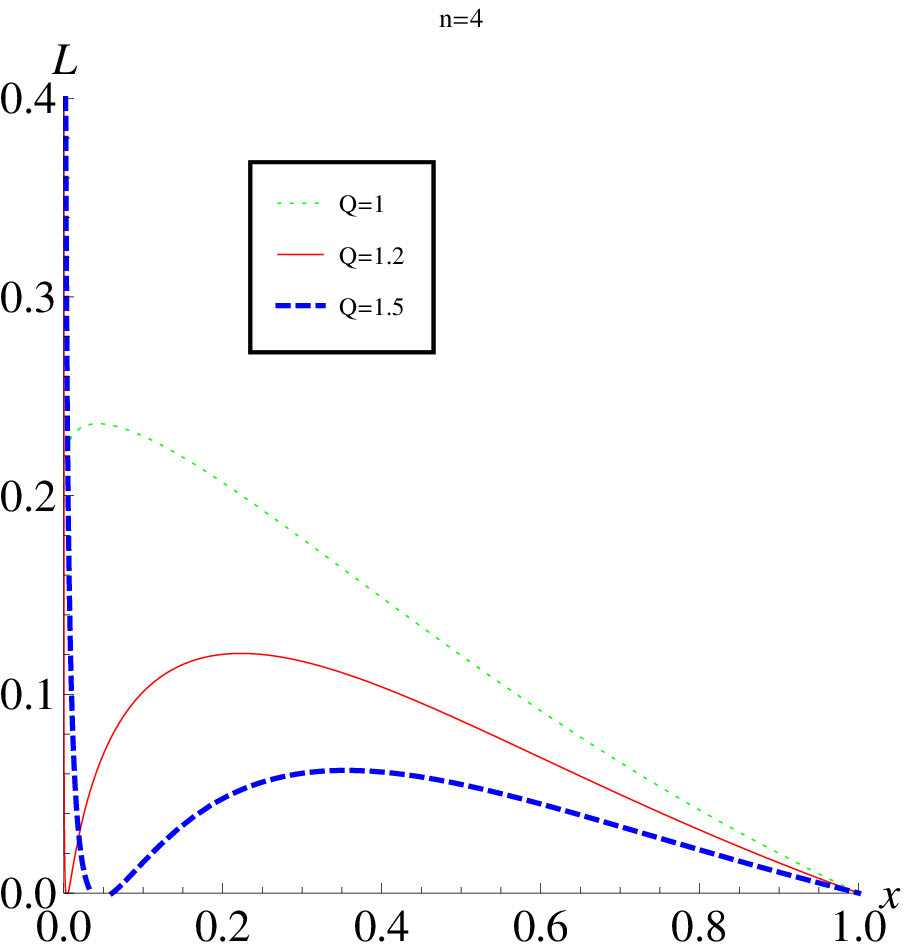}
  \includegraphics[angle=0,width=5cm,keepaspectratio]{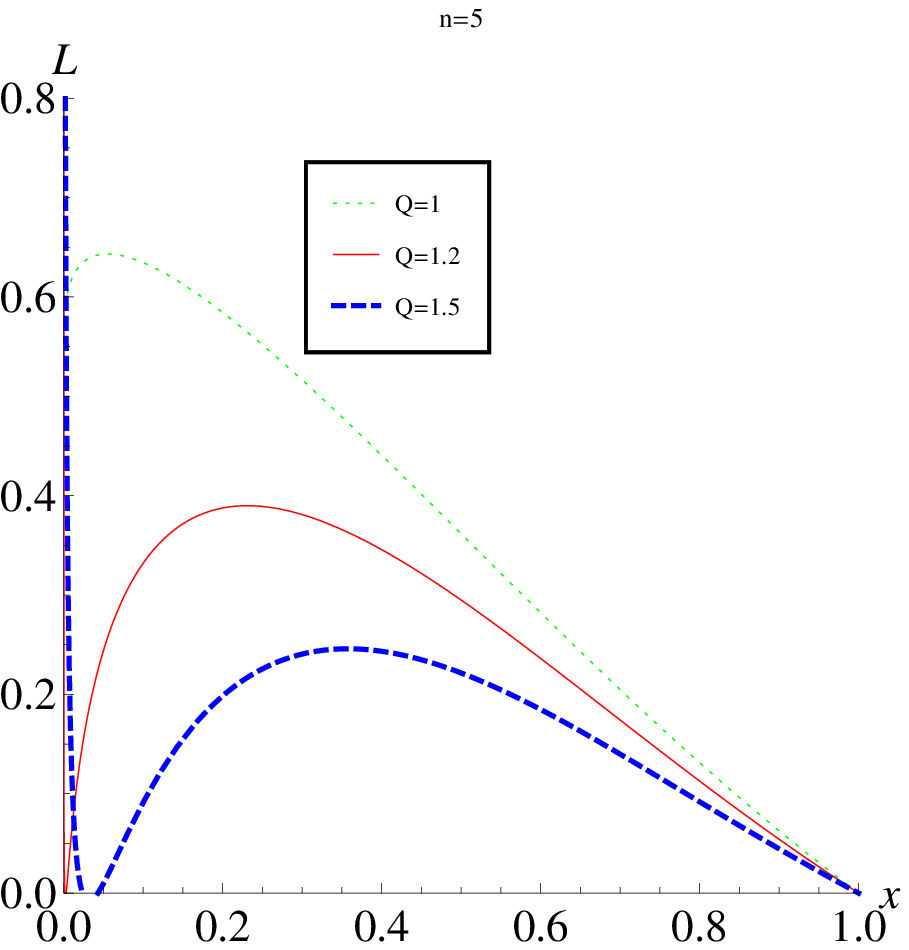}
  \includegraphics[angle=0,width=5cm,keepaspectratio]{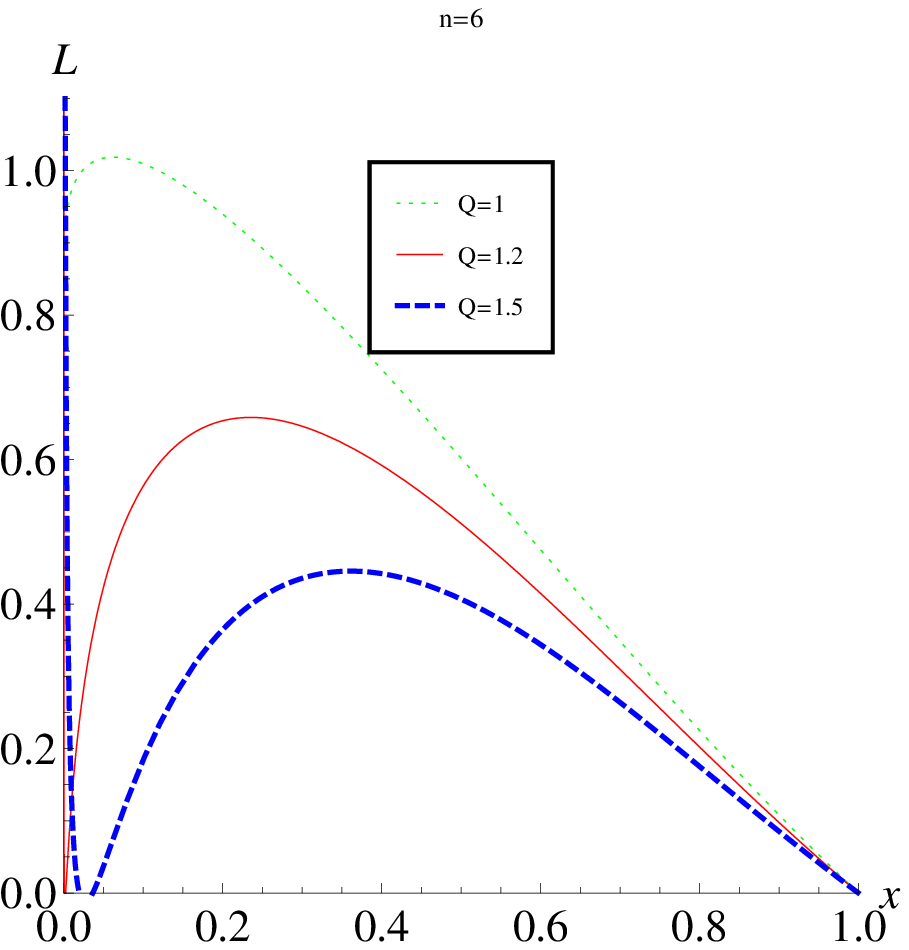}\\
  \caption{\it $L-x$ curves for the topological dilaton black hole in $n$-dimensional
AdS spacetime as $n=4,5,6$. In each diagram, the coupling coefficient $\alpha=\frac{1}{\sqrt{3}}$ and the arbitrary constant $b=1.2$.}\label{L-T0}
\end{figure}

The change rate of latent heat of phase change with temperature for some usual
thermodynamic systems
\begin{equation}
\label{eq46}
\frac{dL}{dT}=C_P^\beta -C_P^\alpha +\frac{L}{T}-\left[ {\left(
{\frac{\partial v^\beta }{\partial T}} \right)_P -\left( {\frac{\partial
v^\alpha }{\partial T}} \right)_P } \right]\frac{L}{v^\beta -v^\alpha },
\end{equation}
where $C_P^\beta$ and $C_P^\alpha$ are molar heat capacity of phase $\beta$
and phase $\alpha $. For $n+1$-dimensional charged topological dilaton AdS
black hole, the change rate of latent heat of phase transition with temperature
can be obtained from Eqs.(\ref{eq45}) and (\ref{eq41}),
\begin{equation}
\label{eq47}
\frac{dL}{dT}=\frac{dL}{dx}\frac{dx}{dT}=\frac{dL}{dx}\frac{1}{G '(x)}.
\end{equation}

\begin{figure}
  \centering
  \includegraphics[angle=0,width=4.7cm,keepaspectratio]{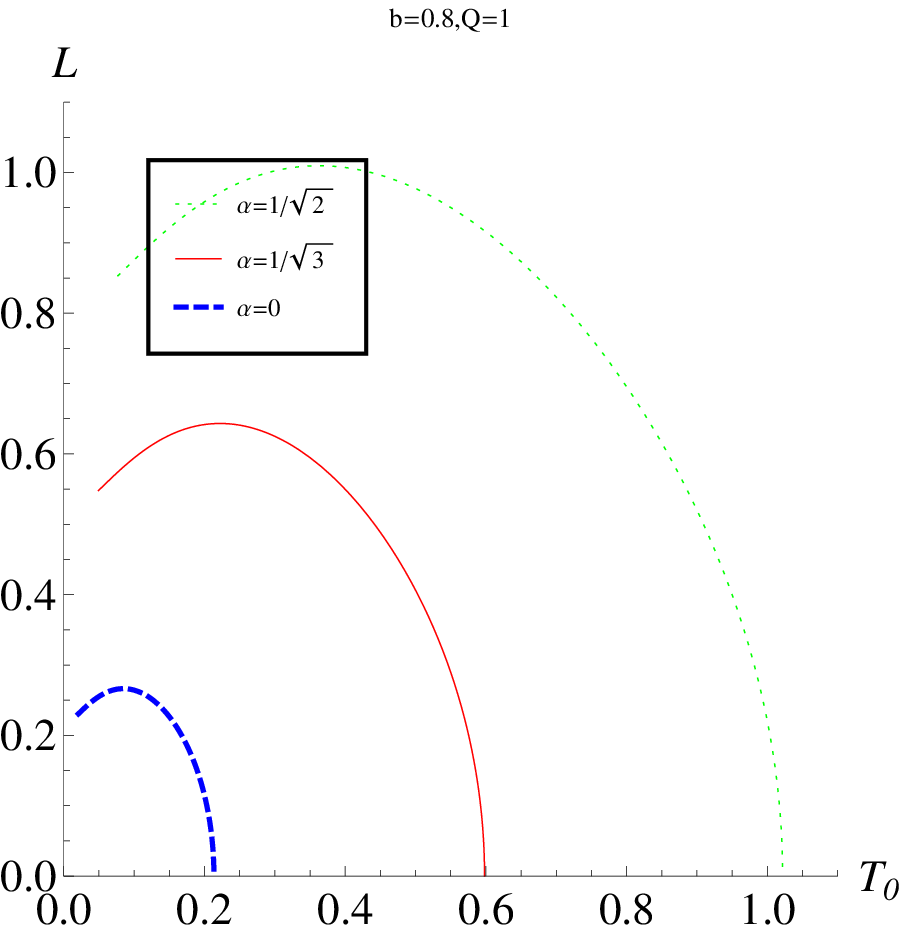}
  \includegraphics[angle=0,width=4.7cm,keepaspectratio]{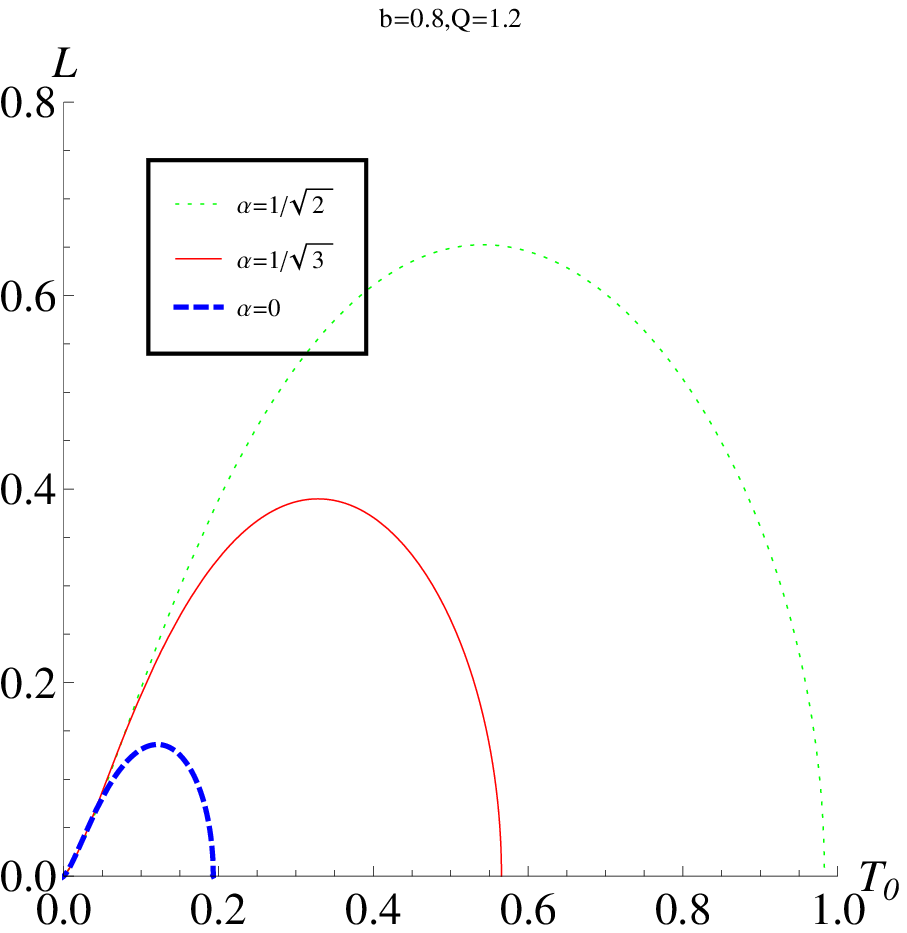}
  \includegraphics[angle=0,width=4.7cm,keepaspectratio]{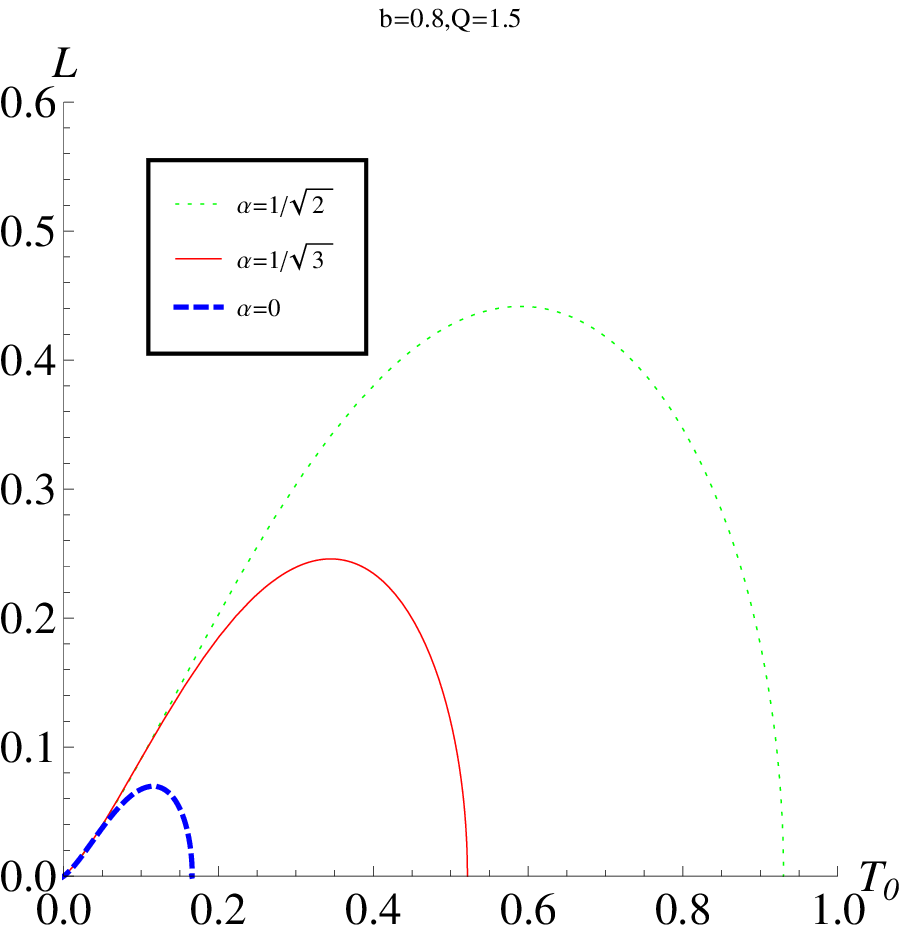}
  \includegraphics[angle=0,width=4.7cm,keepaspectratio]{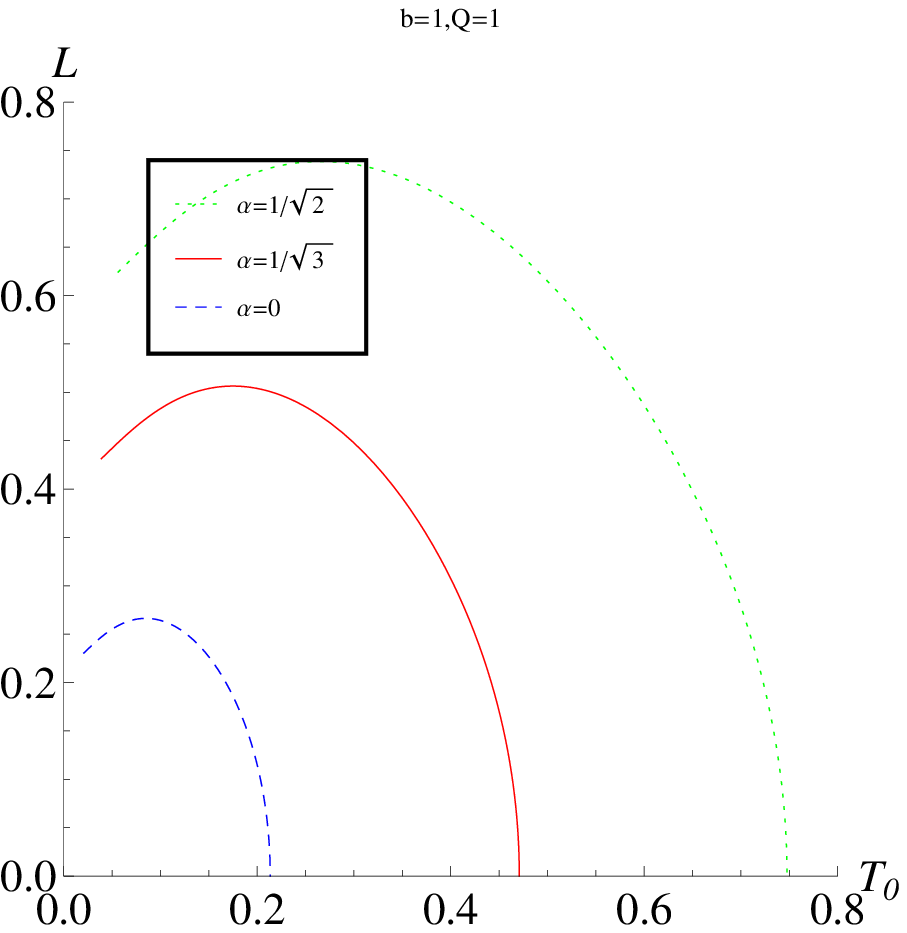}
  \includegraphics[angle=0,width=4.7cm,keepaspectratio]{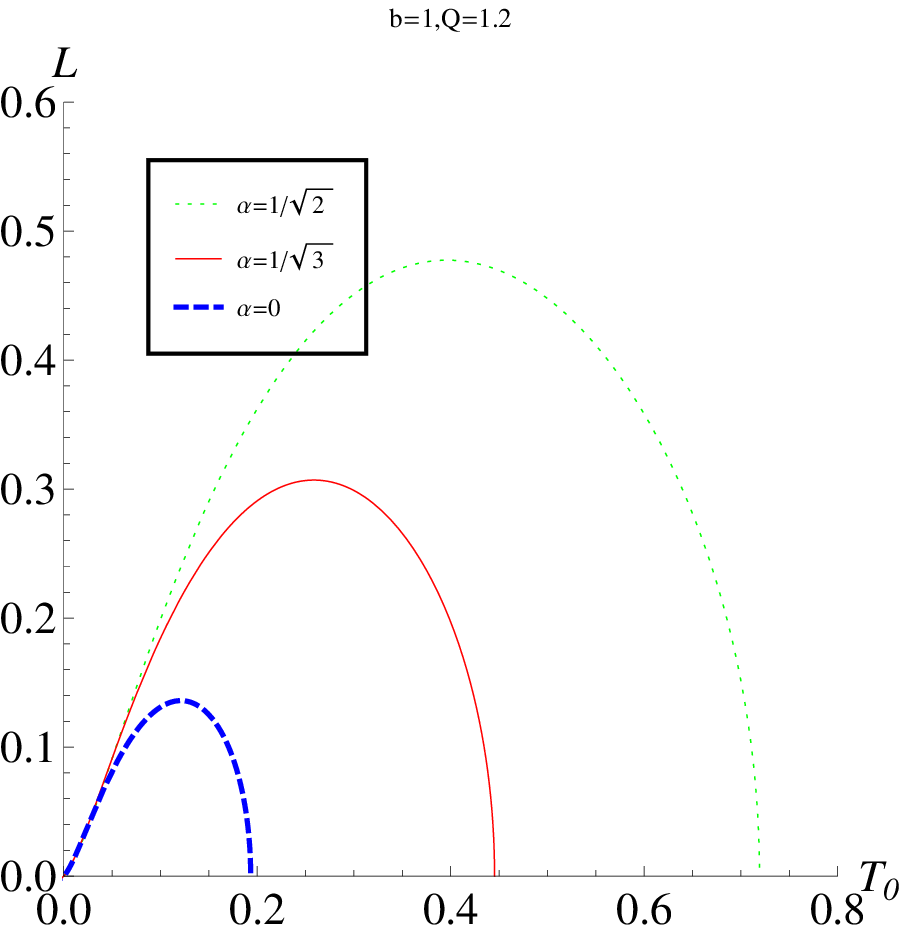}
  \includegraphics[angle=0,width=4.7cm,keepaspectratio]{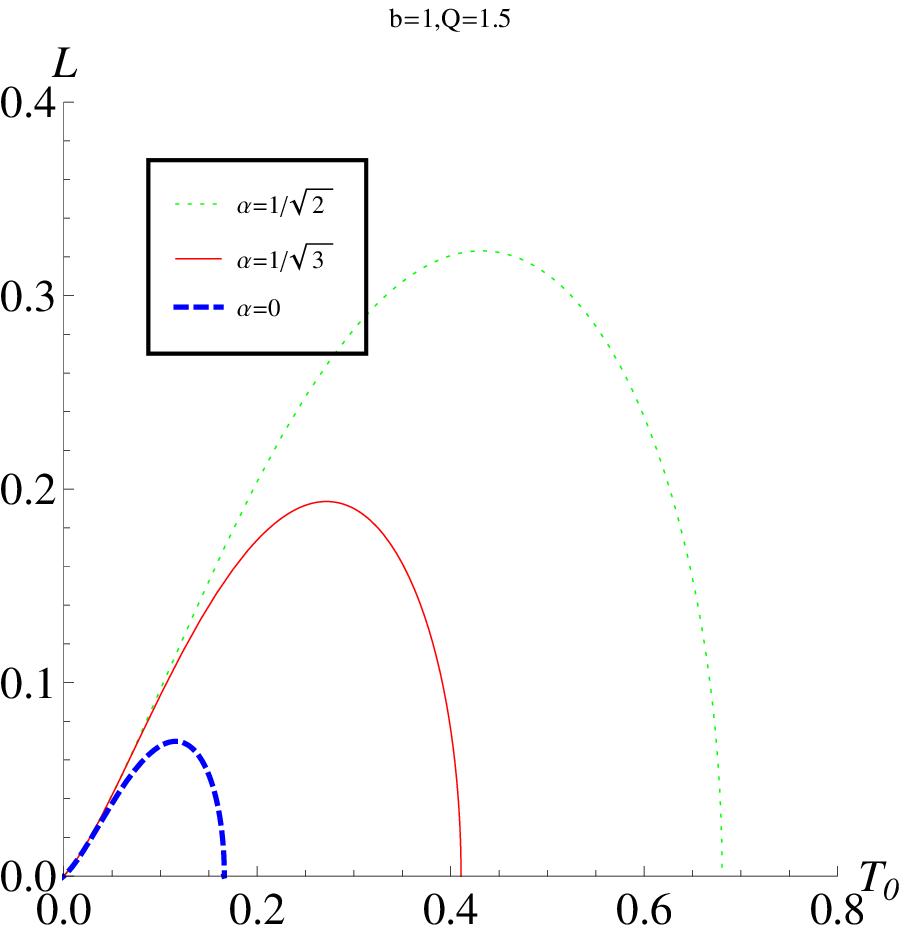}
  \includegraphics[angle=0,width=4.7cm,keepaspectratio]{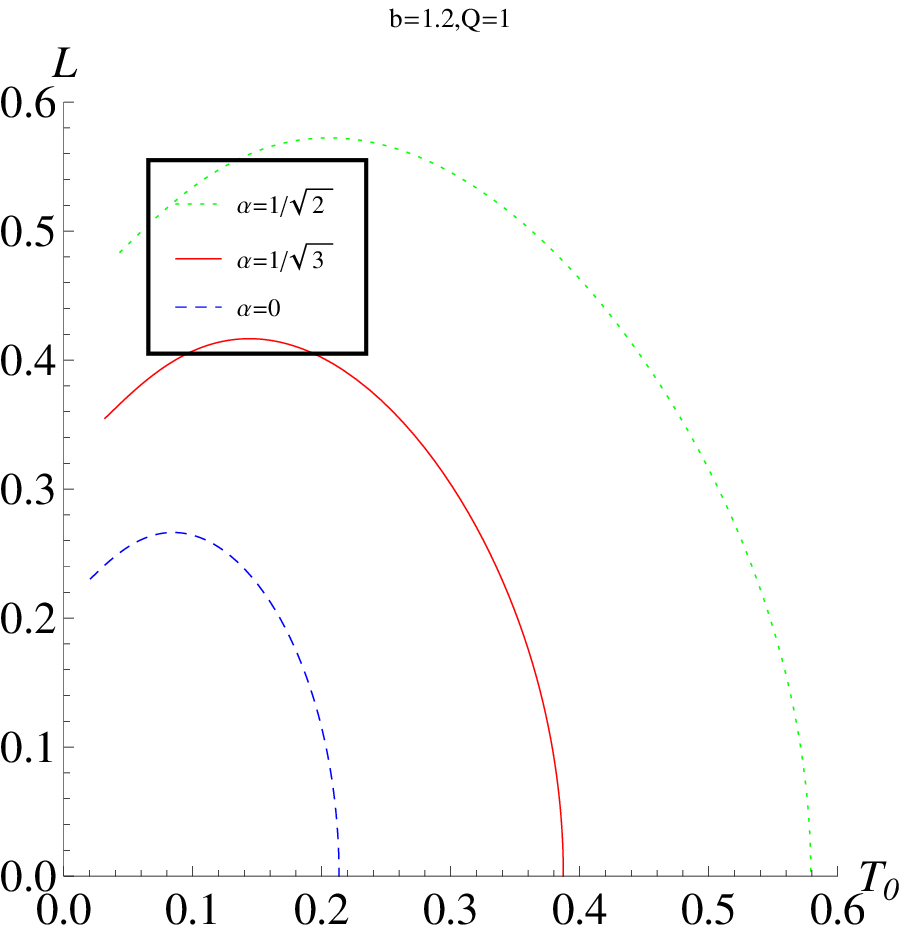}
  \includegraphics[angle=0,width=4.7cm,keepaspectratio]{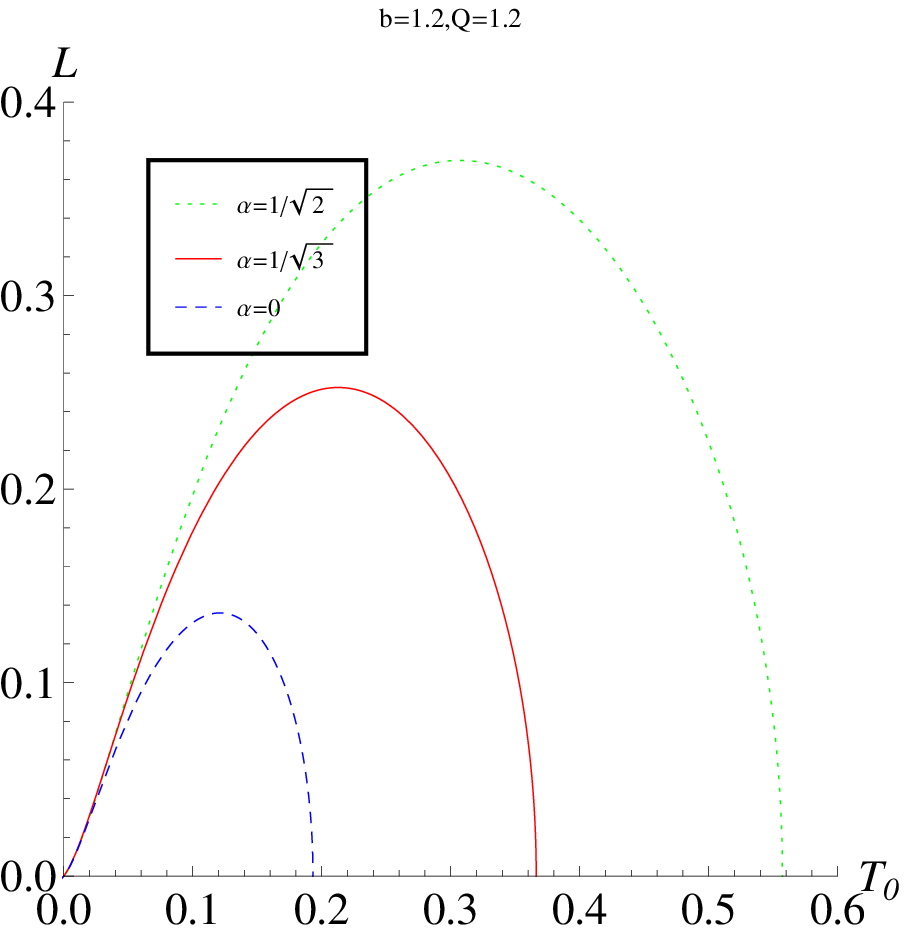}
  \includegraphics[angle=0,width=4.7cm,keepaspectratio]{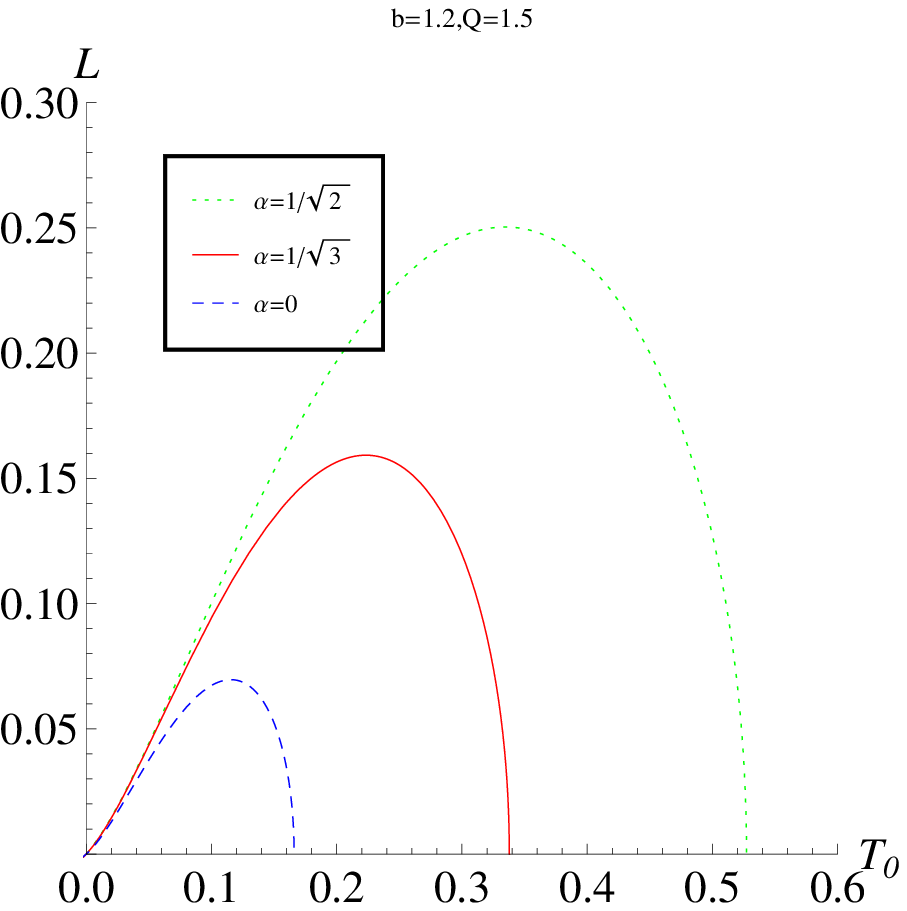}\\
  \caption{\it $L-T$ curves for the topological dilaton black hole in $n$-dimensional
AdS spacetime as $n=5$. In the diagram, the first line with $b=0.8$, the Second line with $b=1.0$, the final line with $b=1.2$ and the first column with
$Q=1.0$, the second column with $Q=1.2$, the final column with $Q=1.5$}\label{L-T0}
\end{figure}
Using Eqs. (\ref{eq45}) and (\ref{eq41}) we plot the latent heat and the temperature
 curves in Fig.5 as the parameter $b$, coupling parameter $\alpha $ and charge
 $Q$ take certain values. From Fig.5 we can see that the effects of temperature $T$ and the coupling parameters $\alpha
$, parameter $b$, and charge $Q$ on latent heat $L$ of phase change. When $T$
increases, the latent heat $L$ is not monotonous but increases firstly and then
decreases to zero as $T\to T_c $. The latent heat $L$ decreases with increasing coupling parameter $\alpha$
as other parameter $b$ and charge $Q$ are fixed. Similarly the latent heat $L$
decreases with increasing charge $Q$ for fixed parameter $b$ and coupling parameter $\alpha$.

\section{Concluding Remarks}

Investigation on the phase transition of the black holes is important and necessary. In
one hand, it is helpful for us to understand the structure and nature of the space time.
In the other hand, it may uncover some phase transitions of the realistic physics in the
conformal field theory according to the AdS/CFT correspondence. The higher-dimensional
charged topological dilaton AdS black hole with a non-linear source is regarded as a
thermodynamic system, and its equation of state has been derived.
But when temperature is below critical temperature,
thermodynamic unstable situation appears on isotherms, and when
temperature reduces to a certain value the negative pressure
emerges, which can be seen in Fig.1 and Fig.2. However, by Maxwell's
equal law we established an phase transition process and the problems can be
resolved. The phase transition process at a defined temperature
happens at a constant pressure, where the system specific
volume changes along with the ratio of the two coexistent phases.
According to Ehrenfest scheme the phase transition belongs to the
first order. We draw the isothermal
phase transition process and depict the boundary of two-phase
coexistence region in Fig.2. In this paper we have plotted the two-phase equilibrium
curves in $P-T$ diagrams, derived the slope of the curves, and
acquired information on latent heat of phase change by Clapeyron equation,
which could create condition for finding some usual thermodynamic systems
similar to black holes in thermodynamic properties and provide theoretical
basis for experimental research on analogous black holes.

In this paper, we have extended the study on the exponential nonlinear electrodynamics by taking into account the
dilaton scalar field in the action. It is important to mention that one can see the influences
of the nonlinear parameter on the thermodynamic properties of the black hole system from Figs. 3 and 5.
For fixed $Q$, $\alpha$, and temperature, the pressure of the two-phase coexistence
decreases as the parameter $b$ decreases. Besides, the latent heat $L$ also decreases
as the parameter $b$ increases. Analyzing these the data, it is found that the phase transition
of black hole thermodynamics become harder if the coupled coefficient $\alpha$ grows up or the
parameter $b$ increases. From Figs.4 one can find that the latent heat $L$ increases
as the spacetime dimension $n$ increases. The latent heat $L$ is more different in the
higher dimension than $4$ dimension spacetime. Meanwhile, we find that the dimension $n$ have an effect
on the simulated phase transition process. The phase transition will be harder when the dimension $n$ is large.

Appropriate theoretical interpretation to the phase structure of the AdS
black hole thermodynamic system can help to know more about black hole thermodynamic
properties, such as entropy, temperature, heat capacity and
so on, of black hole and that is significant for improving
self-consistent thermodynamic theory of black holes. The Clapeyron equation
of usual thermodynamic system agrees well with
experiment result. The thermodynamics of higher order gravity remains to be further explored.
It is quite possible that these objects may also exhibit interesting thermodynamic behaviour by
Maxwell's equal area laws.

\medskip

\appendix

\section{The calculation of the metric function $f(r)$}
\label{A}
The metric function of $f(r)$ is
\[
f(r) = - \frac{(n - 3)(\alpha ^2 + 1)^2b^{ - \gamma }r^\gamma }{(\alpha ^2 -
1)(\alpha ^2 + n - 3)} - \frac{m}{r^{n - 3 - (n - 2)\gamma / 2}} +
\frac{2(2\Lambda + 2\beta ^2)(\alpha ^2 + 1)^2b^\gamma }{(n - 2)(\alpha ^2 -
n + 1)}r^{2 - \gamma }
\]
\[
 + \frac{2\beta q(\alpha ^2 + 1)^4b^{(4 - n)\gamma / 2}}{(n - 2)^2(\alpha ^2
- 1)^2}\left( {\frac{\beta ^2b^{(n - 2)\gamma }}{q^2}}
\right)^{\textstyle{{1 - \gamma } \over {(n - 2)(\gamma - 2)}}}r^{(n -
2)\gamma / 2 - n + 3}
\]

\[
\times \left( {\frac{1 - \alpha ^2}{2n - 4}} \right)^{\textstyle{{2 - 2n +
\gamma n} \over {(\gamma - 2)(2n - 4)}}}\left\{ { - (n - 2)^2(\gamma -
2)^2\left[ {\Gamma \left( {\frac{\alpha ^2 + 3n - 7}{2n - 4},\frac{1 -
\alpha }{2n - 4}L_W (\eta )} \right) - \Gamma \left( {\frac{\alpha ^2 + 3n -
7}{2n - 4}} \right)} \right]} \right.
\]
\begin{equation}
\label{eq4}
 + (\gamma - 1)\left. {^2\left[ {\Gamma \left( {\frac{\alpha ^2 - n + 1}{2n
- 4},\frac{1 - \alpha ^2}{2n - 4}L_W (\eta )} \right) - \Gamma \left(
{\frac{\alpha ^2 - n + 1}{2n - 4}} \right)} \right]} \right\} ,
\end{equation}

\begin{equation}
\label{eq5}
\Phi (r) = \frac{(n - 2)\alpha }{2(\alpha ^2 + 1)}\ln \left( {\frac{b}{r}}
\right)
\end{equation}
with $b$ is an arbitrary constant, $\gamma =\alpha ^2/(\alpha ^2+1)$, and
\begin{equation}
\label{eq6}
\eta\equiv \frac{q^2 b^{(2-n)\gamma}}{\beta^2 r^{(n-2)(2-\gamma)}}
\end{equation}
$m$ appears as an integration constant and is related
to the mass of the black hole, $q$ is an integration constant which is
related to the electric charge of the black hole. The electric charge is
\begin{equation}
\label{eq7}
Q = \frac{q\omega _{n - 2} }{4\pi },
\end{equation}
where $\omega _{n - 2}$ represents the volume of constant curvature
hypersurface described by $d\Omega _{n - 2}^2 $, $L_W (x)$ is
the Lambert function which satisfies the identity~\cite{Thar}
\begin{equation}
\label{eq8}
L_W (x)e^{L_W (x)} = x,
\end{equation}
and has the following series expansion
\begin{equation}
\label{eq9}
L_W (x) = x - x^2 + \frac{3}{2}x^3 - \frac{8}{3}x^4 + \cdots .
\end{equation}
$\Gamma (a,z)$ and $\Gamma (a)$ are gamma functions and they are related to each
other via
\begin{equation}
\label{eq10}
\Gamma (a,z) = \Gamma (a) - \frac{z^a}{a}F(a,1 + a, - z),
\end{equation}
where $F(a,a,z)$ is hypergeometic function.

Using the fact that $L_W (x)$ has a convergent series expansion for $\left|
x \right| < 1$ as given in (\ref{eq9}), we can expand (\ref{eq4}) for large $\beta $.
The result is~\cite{Sheykhi1}
\[
f(r) = - \frac{(n - 3)(\alpha ^2 + 1)^2b^{ - \gamma }r^\gamma }{(\alpha ^2 -
1)(\alpha ^2 + n - 3)} - \frac{m}{r^{n - 3 - (n - 2)\gamma / 2}} +
\frac{2\Lambda (\alpha ^2 + 1)^2b^\gamma }{(n - 2)(\alpha ^2 - n + 1)}r^{2 -
\gamma }
\]
\begin{equation}
\label{eq11}
 + \frac{2q^2(\alpha ^2 + 1)^2b^{ - (n - 3)\gamma }}{(n - 2)(\alpha ^2 + n -
3)r^{ - (n - 3)\gamma + 2n - 6}} - \frac{q^4(\alpha ^2 + 1)^2b^{ - (2n -
5)\gamma }}{2\beta ^2(n - 2)(\alpha ^2 + 3n - 7)r^{(2n - 5)(2 - \gamma )}} +
o\left( {\frac{1}{\beta ^4}} \right).
\end{equation}

\begin{acknowledgments}\vskip -4mm
We would like to thank Dr Meng-Sen Ma and Yu-Bo Ma for their indispensable discussions and comments.This work was supported by the Young Scientists Fund of the National Natural Science Foundation of China (Grant No.11205097), in part by the National Natural Science Foundation of China (Grant No.11475108), Supported by Program for the Innovative Talents of Higher Learning Institutions of Shanxi, the Natural Science Foundation of Shanxi Province,China(Grant No.201601D102004) and the Natural Science Foundation for Young Scientists of Shanxi Province,China (Grant No.2012021003-4), the Natural Science Foundation of Datong city(Grant No.20150110).

\end{acknowledgments}

\end{document}